\documentclass[twocolumn,aps]{revtex4}
\usepackage{graphicx}

\newcommand{\mvec}[1]{{\mbox{{\boldmath$#1$}}}}
\newcommand{\be}{\begin{equation}}
\newcommand{\ee}{\end{equation}}
\newcommand{\bea}{\begin{eqnarray}}
\newcommand{\eea}{\end{eqnarray}}

\begin{document}

\preprint{BAS-04}
\title{{\LARGE \textbf{Path integral derivation of Bloch-Redfield equations
for a qubit weakly coupled to a heat bath: Application to nonadiabatic
transitions}}}
\author{Nicholas~W.~Watkins}
\affiliation{On Leave at Center for Space Research, Massachusetts Institute of
Technology, Cambridge, Mass., USA}
\altaffiliation{Permanent address: Physical Sciences Division, British Antarctic Survey
(NERC), High Cross, Madingley Road, Cambridge, UK}
\email{NWW@bas.ac.uk}
\author{David~Waxman}
\affiliation{School of Life Sciences, University of Sussex, Brighton, Sussex, UK}
\date{\today}

\begin{abstract}
Quantum information processing has greatly increased interest in the
phenomenon of environmentally-induced decoherence. The spin boson model is
widely used to study the interaction between a spin-modelling a quantum
particle moving in a double well potential-and its environment-modelled by a
heat bath of harmonic oscillators. This paper extends a previous analysis of
the static spin boson study to the driven spin boson case, with the
derivation of an exact integro-differential equation for the time evolution
of the propagator of the reduced spin density matrix. This is the first main
result. By specializing to weak damping we then obtain the next result, a
set of Bloch-Redfield equations for the equilibrium fixed spin initial
condition. Finally we show that these equations can be used to solve the
classic dissipative Landau-Zener problem and illustrate these solutions for
the weak damping case. The effect of dissipation is seen to be minimised as
the speed of passage is increased, implying that qubits need to be switched
as fast as possible.
\end{abstract}

\maketitle



\section{Introduction}

\label{Introduction} A new era of quantum information processing
experiments is beginning (e.g.
\cite{NakamuraNature1999,FriedmanNature2000,vanderWalScience2000})
with the consequence that battling decoherence (the destruction by
environmental degrees of freedom of the phase coherence between
superposed quantum states) has become a crucial task for designing
the fault-tolerant quantum computer \cite{Preskill1999}. The
engineering significance now attaching to decoherence gives new
urgency to an already fundamental theoretical question (e.g.
\cite{Leggett1987,Grifoni1998,TakagiBook}): what is the effect of
an environmental heat bath on a quantum two-level system (qubit)?

As well as its importance for the experimental realization of a qubit \cite%
{Shytov2000,IoffeNature2002,Shytov2003}, the current interest in this
problem is due to at least two other factors. One is its general importance
as a testbed for competing conceptions of dissipative quantum mechanics \cite%
{WeissBook,BreuerBook}. A second is the bridge it offers between quantum and
classical stochastic systems, for example as a quantum rather than classical
two level stochastic resonance system (e.g. \cite{stocr,Grifoni1998}).

A particularly well-studied framework for the problem is the spin-boson
Hamiltonian (e.g. \cite{Leggett1987, Grifoni1998}) in which the two level
system is modelled by a spin, the environmental heat bath by quantum
harmonic oscillators, and the spin is coupled to each bath oscillator
independently. This Hamiltonian is easily adaptable to the driven spin-boson
case where the two level system is subjected to an external force. In the
strong coupling and high temperature bath limits much progress has been
made, especially for the case of harmonic driving which defines the quantum
stochastic resonance problem \cite{stocr,Grifoni1998}. Hartmann \textit{et
al.} \cite{Hartmann2000} have, however, recently emphasized that weak
coupling and low temperatures remain much less well explored, and that
although formal solutions for the resulting spin dynamics do now exist, they
still in practise have to be used perturbatively.

Hartmann \textit{et al.} \cite{Hartmann2000} sought to make progress by
pointing out similarities of the driven spin boson problem to counterparts
in quantum optics and solid state physics where Bloch-Redfield equations are
used. They exhibited a set of Bloch-Redfield equations (their Eqn. 4)
obtained by projection operator methods.

In this Paper we show that an earlier path integral-based derivation of a
similar set of equations \cite{WaxJPC1985} for the time independent
spin-boson Hamiltonian can be extended to the driven case (see also \cite%
{Watkins1989}). This allows us to derive (section~\ref{Propagator}) an \emph{%
exact} integro-differential evolution equation for the propagator
of the reduced density matrix, and is the first of of three
principal results of the Paper. We then
(section~\ref{BlochRedfield}) make the weak coupling assumption to
give an independent derivation of a set of Bloch-Redfield
equations for the driven spin boson case; the second main result
of the
Paper. We note in passing that this refutes the assertion by Zhang \cite%
{Zhang1988} that the methodology used in \cite{WaxJPC1985} was unsuitable
for time dependent problems.

We choose an initial condition in which the environment
equilibrates about a fixed spin, which is subsequently released at
time $t_{0}$. We find, in consequence, a difference from the
Bloch-Redfield equations of \cite{Hartmann2000}, but only in the
terms arising from dissipation. This difference disappears if we
follow previous authors in assuming spin and bath to be completely
uncoupled at $t_{0}$. In either case our terms due to fluctuation
are identical with those seen by \cite{Hartmann2000}.

The main application we study for our Bloch-Redfield equations is the
dissipative version of the classic Landau-Zener-Stuckelberg (LZS)
nonadiabatic level crossing problem (e.g. \cite%
{A&R1989,WaxmanAP,NakamuraGreenBook,Grifoni1998}). This describes
the situation where the energy levels of a quantum two-level
system are brought together over time by an external force,
enabling transitions to take place. Examples of this ubiquitous
problem are found in atomic collisions, chemical reaction
kinetics, biophysics, solar neutrino oscillations and NMR (see Ao
and Rammer \cite{A&R1989} for references and additional examples).
In the present case we will take the LZS Hamiltonian with a fixed
tunnelling term and linearly time-varying bias, and study the
effects of an environment consisting of harmonic oscillators. The
initial study of of the dissipative LZS problem by Ao and Rammer
\cite{A&R1989} has since been
supplemented by other analytic and computational methods (e.g. \cite%
{NakamuraGreenBook,KayanumaNakayama,Shytov2000}). In one case \cite%
{KayanumaNakayama} these benefit from the recent feasibility of
computationally following $\approx $ 120000 basis states; but they
have all given results for the influence of the bath, for
intermediate speeds of
passage, which disagree with those of \cite{A&R1989}. Here (\ref%
{BlochRedfield}) we specialize the Bloch-Redfield equations we
have derived to the LZS problem in the ohmic weak damping case,
which greatly simplifies the dissipation term (as also found by
\cite{Zhang1988}). These equations for the dissipative LZS problem
may  aid study of topical issues such as the merits of different
models of the noise power spectral density $J(\omega )$ of the
heat bath \cite{Shytov2000} or of alternative functional forms for
the time dependent bias in the LZS problem (c.f. Chapter 6 of
\cite{NakamuraRedBook}).

By scaling the Bloch-Redfield equations to the characteristic
tunnelling time, we show that dissipation is significant even at
zero temperature, for finite speeds of passage, agreeing with
newer authors, but that its impact is inversely proportional to
the passage speed. This implies the intuitively reasonable
conclusion that qubits need to be switched as fast as possible to
minimise dissipation effects. We illustrate our findings
numerically, in the zero temperature case where the fluctuating
force terms greatly simplify in addition to the more general ohmic
simplification of the dissipation term. Damping is clearly seen,
reminiscent of that seen for the high temperature case in
\cite{WaxmanAP}

\section{Exact integro-differential evolution equation for spin propagator
of driven spin-boson Hamiltonian with equilibrium fixed spin initial
condition}

\label{Propagator}

First, in subsection~\ref{DissLZ} we describe the driven spin-boson system,
defining the Hamiltonian and our notation (based on \cite%
{Leggett1987,WaxJPC1985,ChangWaxman}). In subsection~\ref{InfFunc} we write
down the reduced spin density matrix as a path integral, to which
(subsection ~\ref{EFS}) we apply the equilibrium fixed spin initial
condition. We can then (subsection~\ref{EFSIF} and~\ref{EFSIF2}) integrate
out the bath variables to find the effect of the bath on the spin by means
of the Feyman-Vernon influence functional for this initial condition. This
enables us to generalize the results of Waxman \citep{WaxJPC1985} to the
driven spin boson case, arriving at our first main result (~\ref{eq:Kdif}),
a general time evolution integro-differential equation for the propagator $K$
of the reduced density matrix $\tilde{\rho}$.

In this section we make only two main assumptions in addition to those of
the spin-boson model itself: i) the widely-employed (e.g. \cite{A&R1989})
choice of a factorizing initial density matrix and ii) use of an
``equilibrium fixed spin" initial condition where the \textit{spin is
assumed to be fixed ``up" and in equilibrium with the bath at $t_0$}. To aid
subsequent comparison with Hartmann \textit{et al.} \cite{Hartmann2000} we
also give the equivalent influence function for the case when spin and bath
are uncoupled at $t_0$.

\subsection{The Driven Spin Boson problem}

\label{DissLZ}

The driven spin-boson Hamiltonian $H$ is:
\begin{equation}
H=H_{S}+H_{I}+H_{B}  \label{eq:spinbosonH}
\end{equation}%
where
\begin{equation}
H_{S}=-\frac{\hbar \Delta }{2}\sigma _{x}+\frac{\hbar \epsilon (t)}{2}\sigma
_{z}
\end{equation}%
\begin{equation}
H_{I}=\sum_{\alpha }\frac{q_{0}}{2}\sigma _{z}c_{\alpha }x_{\alpha }
\end{equation}%
\begin{equation}
H_{B}=\sum_{\alpha }\left( \frac{p_{\alpha }^{2}}{2m_{\alpha }}+\frac{%
m_{\alpha }}{2}x_{\alpha }^{2}\omega _{\alpha }^{2}\right)
\end{equation}%
where $\sigma _{i}$ with $i=x,y,z$ are Pauli spin operators; $-\Delta $ and $%
\epsilon (t)$ are angular frequencies corresponding to off-diagonal
(tunnelling) and on-diagonal (bias) spin matrix elements respectively; and
the heat bath is represented by a set of harmonic oscillators of mass $%
m_{\alpha }$, angular frequency $\omega _{\alpha }$, momentum $p_{\alpha }$
and position coordinate $x_{\alpha }$. The oscillators are coupled
independently to the spin co-ordinate with strength measured by the set $%
\{c_{\alpha }\}$ while $q_{0}$ measures the distance between the left and
right potential wells.

In contrast to most previous work (e.g. \cite{WaxJPC1985}) we will
use an initial condition where at times $-\infty $ to $t_{0}$ the
spin has been held fixed and the heat bath has equilibrated around
it; a choice which requires explanation. For reasons of
mathematical simplicity we want a factorizing initial full density
matrix. Given this, we have selected what we believe to be the
boundary condition that avoids unnatural transients associated
with the time evolution of the complete system after $t_{0}$. Any
other choice for the bath's initial density matrix would be
expected to lead to transients because at time $t_{0}$ both the
bath and the spin would have dynamics driven by the interaction
between them. Such an initial condition can be viewed as the
result of having $\epsilon (t)$ very large and negative for times
$t<t_{0}$.

We  thus take
\begin{equation}
H(t=t_{0})=\sum_{\alpha }\left[ \frac{p_{\alpha }^{2}}{2m_{\alpha }}+\frac{%
m_{\alpha }\omega _{\alpha }^{2}}{2}\left( x_{\alpha }+\frac{q_{0}c_{\alpha }%
}{2m_{\alpha }\omega _{\alpha }^{2}}\right) ^{2}\right]  \label{eq:freeH}
\end{equation}%
which incorporates the initial condition for the combined interaction and
bath Hamiltonian at time $t_{0}$. This results from~(\ref{eq:spinbosonH})
with the eigenvalue of $\sigma _{z}$ set at $+1$. The extra term relative
to~(\ref{eq:spinbosonH}), namely
\[
\sum_{\alpha }\left( \frac{q_{0}c_{\alpha }}{2m_{\alpha }\omega _{\alpha
}^{2}}\right) ^{2}\frac{m_{\alpha }\omega _{\alpha }^{2}}{2}
\]%
cancels with its equivalent in the partition function (the denominator of (%
\ref{eq:therm}).

\subsection{The density matrix as a path integral}

\label{InfFunc}

We follow the method of \cite{WaxJPC1985} but the driven spin-boson
Hamiltonian~(\ref{eq:spinbosonH}) replaces the static spin-boson Hamiltonian
given as his equation (1). We start by expressing the density matrix of the
combined system as a double path integral. The density operator $\rho $
obeys $i\hbar \partial \rho /\partial t=[H,\rho ]$, so its time evolution is
given by $\rho (t)=U(t,t_{0})\rho (t_{0})U^{-1}(t,t_{0})$ with $U$ a unitary
time evolution operator.

Hence with $|{\mbox{{\boldmath$x$}}}\sigma >$ an eigenstate of the
oscillators' coordinate operators ${\mbox{{\boldmath$x$}}}$ and the spin
operator $\sigma _{z}$ we have
\begin{widetext}
\begin{eqnarray}
  < \mvec{x}_1 \sigma_1 | \rho (t) | \mvec{x}_2 \sigma_2 >
& = &
  \rho(\mvec{x}_{1} \sigma_1 , \mvec{x}_{2} \sigma_2 ; t ) \nonumber \\
& =&
    \sum_{\sigma_{3}}\sum_{\sigma_{4}}\int\!d\mvec{x_{3}}
                      \int\!d\mvec{x_{4}}\,
    <\mvec{x}_{1}\sigma_{1}|U(t,t_0)|\mvec{x}_{3}\sigma_3>
\nonumber \\
& &
    \times<\mvec{x}_{3}\sigma_{3}|\rho(t_0)|\mvec{x}_{4}\sigma_4>
    <\mvec{x}_{4}\sigma_{4}|U^{-1}(t,t_0)|\mvec{x}_{2}\sigma_2>
\end{eqnarray}
\end{widetext}where we have inserted two complete sets of states and
boldface type distinguishes a state of all $N$ oscillators (${%
\mbox{{\boldmath$x$}}}\equiv \{x_{\alpha }\}$). We now assume that the
initial density matrix factors into an oscillator dependent part and a spin
dependent part. We are only interested in the behaviour of the spin so work
with the reduced density matrix, obtained by integrating over the bath
variables. We then define the reduced density matrix $\tilde{\rho}$ by
tracing over the bath variables, and normalizing to the free oscillator
partition function. We have
\begin{equation}
\tilde{\rho}(\sigma _{1}\sigma _{2}t)=\sum_{\sigma _{3}}\sum_{\sigma
_{4}}K(\sigma _{1}\sigma _{2}t|\sigma _{3}\sigma _{4}t_{0})\tilde{\rho}%
(\sigma _{3}\sigma _{4}t_{0}).  \label{eq:reddm2}
\end{equation}%
Our initial problem is to determine the propagator $K$:
\begin{eqnarray}
K(\sigma _{1}\sigma _{2}t|\sigma _{3}\sigma _{4}t_{0}) &=&\int \!d{%
\mbox{{\boldmath$x$}}}^{\prime }\int \!d{\mbox{{\boldmath$x$}}}_{3}\int \!d{%
\mbox{{\boldmath$x$}}}_{4}\langle {\mbox{{\boldmath$x$}}}^{\prime }\sigma
_{1}|U(t,t_{0})|{\mbox{{\boldmath$x$}}}_{3}\sigma _{3}\rangle  \nonumber \\
&&\times \rho ({\mbox{{\boldmath$x$}}}_{3}{\mbox{{\boldmath$x$}}}%
_{4}t_{0})\langle {\mbox{{\boldmath$x$}}}_{4}\sigma _{4}|U^{-1}(t,t_{0})|{%
\mbox{{\boldmath$x$}}}^{\prime }\sigma _{2}\rangle
\end{eqnarray}%
the effective time evolution operator for the reduced density matrix.

We now write the matrix element of the time evolution operator $U(t,t_{0})$
(the forward propagator) as a path integral
\begin{equation}
\langle {\mbox{{\boldmath$x$}}}_{1}\sigma _{1}|U(t,t_{0})|{%
\mbox{{\boldmath$x$}}}_{3}\sigma _{3}\rangle =\int_{{\mbox{{\boldmath$x$}}}%
_{3}t_{0}}^{{\mbox{{\boldmath$x$}}}_{1}t}\!d[x]\int_{\sigma
_{3}t_{0}}^{\sigma _{1}t}\!d[\sigma ]\exp \left( \frac{i}{\hbar }S[\sigma
,x]\right)
\end{equation}%
with
\begin{equation}
S[\sigma ,x]=S_{S}[\sigma ]+S_{I}[\sigma ,x]+S_{B}[x]
\end{equation}%
where $S_{S},S_{I},S_{B}$ are the actions corresponding to the spin,
interaction and bath Hamiltonian operators $H_{S},H_{I}$ and $H_{B}$
respectively. The notation
\begin{equation}
\int_{{\mbox{{\boldmath$x$}}}_{3}t_{0}}^{{\mbox{{\boldmath$x$}}}_{1}t}\,d[x]
\end{equation}%
means the sum over all paths beginning at ${\mbox{{\boldmath$x$}}}_{3}$ at
time $t_{0}$ and ending at ${\mbox{{\boldmath$x$}}}_{1}$ at time $t$, and
the spin path integral is explained more fully below.

Combining the above with a similar expression for the backward propagator
and with the initial bath density matrix we obtain

\begin{equation}
K(\sigma _{1}\sigma _{2}t|\sigma _{3}\sigma _{4}t_{0})=\int_{\sigma
_{3}t_{0}}^{\sigma _{1}t}\!d[\sigma ]\int_{\sigma _{4}t_{0}}^{\sigma
_{2}t}\!d[\nu ]A_{n}[\sigma ]A_{m}^{\ast }[\nu ]\,F[\sigma ,\nu ]
\label{eq:K}
\end{equation}%
where
\begin{widetext}
\begin{eqnarray}
 F[\sigma ,\nu ]
&=& \int\!d\mvec{x}'\int\!d\mvec{x}_{3}\int\!d\mvec{x}_{4}
 \int_{\mvec{x}_{4} t_0}^{\mvec{x}' t}\!d[y]\int_{\mvec{x}_{3} t_0}^
 {\mvec{x}' t}\!d[x]\nonumber\\
& & \times e^{\frac{i}{\hbar}(S_{I}[\sigma ,x]-S_{I}[\nu
,y]+S_{B}[x]-S_{B}[y])} \,\rho (\mvec{x}_{3} \mvec{x}_{4}
t_0)\label{eq:if}
\end{eqnarray}
\end{widetext} is the influence functional containing all effects of the
bath on the spin system.%
\begin{eqnarray}
\int d[\sigma ]\exp \left( \frac{i}{\hbar }S[\sigma ]\right)
&=&\sum_{n}\int_{t_{0}}^{t}\!dt_{n}\int_{t_{0}}^{t_{n}}\!dt_{n-1}\ldots
\int_{t_{0}}^{t_{2}}\!dt_{1}  \nonumber \\
&&\times \left( \frac{i\Delta }{2}\right) ^{n}\exp \left( -\frac{i}{2}%
\int_{t_{0}}^{t}\!du\epsilon (u)\sigma (u)\right)  \label{eq:spinpi}
\end{eqnarray}%
with the sum[[,]] running over all numbers of flips (points at which $\sigma
$ goes from $-1$ to $+1$ or vice versa) consistent with the initial
conditions, defines both the measure $d[\sigma ]$ and the amplitude $%
A_{n}[\sigma ]$ associated with a path $[\sigma ]$ containing $n$ spin flips.

\subsection{Application of equilibrium fixed spin initial condition}

\label{EFS}

We now calculate the influence functional and so have to specify $\rho ({%
\mbox{{\boldmath$x$}}}_{3} {\mbox{{\boldmath$x$}}}_{4} t_{0} )$, which
corresponds physically to the description of the oscillator bath at time $%
t_{0}$. We assume thermal equilibrium so
\begin{equation}
\rho ({\mbox{{\boldmath$x$}}}_{3} {\mbox{{\boldmath$x$}}}_{4} t_{0} )=<{%
\mbox{{\boldmath$x$}}}_{3} |\frac{e^{-\beta H_{0}}} {Z_{0}} |{%
\mbox{{\boldmath$x$}}}_{4} >  \label{eq:therm}
\end{equation}
in which $H_{0}$ denotes $H(t=t_0)$. The partition function is given by $%
Z_{0}=\mathrm{Tr} e^{-\beta H_{0}}$ and the inverse temperature by $\beta =
\frac {1}{k_B T} $.

We find that the path integral part of the influence functional~(\ref{eq:if}%
) factors into a product of terms like
\begin{widetext}
\begin{equation}
\int_{x_{4}t_{0}}^{x't}d[y]\int_{x_{3}t_{0}}^{x't}d[x] <
x_{3}|\frac{e^{-\beta H_{\alpha 0}}}{Z_{0}}|x_{4}> \exp
\frac{i}{\hbar}(S_{\alpha I}[\sigma ,x]-S_{\alpha I}[\nu
,y]+S_{\alpha B} [x]-S_{\alpha B}[y])
\end{equation}
\end{widetext} one for each [[]] oscillator, $x_{3}$ and $x_{4}$ are the
initial states for the forward and backward paths. Noting that the (single
particle) bath and interaction actions are
\begin{equation}
S_{\alpha I}[\sigma ,x]+S_{\alpha B}[x]=\int_{t_{0}}^{t}du\left( \frac{%
m_{\alpha }\dot{x}_{\alpha }^{2}}{2}-\frac{m_{\alpha }\omega _{\alpha
}^{2}x_{\alpha }^{2}}{2}-\frac{c_{\alpha }x_{\alpha }q_{0}\sigma }{2}\right)
\end{equation}%
we first evaluate the forward propagator for the bath variables, a standard
result \cite{SchulmanBook}:
\begin{widetext}
\begin{eqnarray}
\int_{x_{3}t_{0}}^{x't}d[x]\exp\frac{i}{\hbar}(S_{\alpha I}[\sigma
,x] +
                                                  S_{\alpha B}[x])
& = & \left(\frac{m_{\alpha}\omega_{\alpha}}{2\pi
i\hbar\sin\omega_{\alpha }(t-t_{0})}
\right)^{\frac{1}{2}}\nonumber \\
& &
\times\exp\left(\frac{im_{\alpha}\omega_{\alpha}}{2\hbar\sin\omega_{\alpha}
(t-t_{0})}B\right)
\end{eqnarray}
\end{widetext}with 
\begin{eqnarray}
B &=&\left( x_{3}^{2}+(x^{\prime })^{2}\right) \cos \omega _{\alpha
}(t-t_{0})-2x^{\prime }x_{3}  \nonumber \\
&&-x^{\prime }f_{\sigma }(t-t_{0})+x_{3}g_{\sigma }(t-t_{0})  \nonumber \\
&&-\frac{2c_{\alpha }^{2}}{m_{\alpha }^{2}\omega _{\alpha }^{2}}\left( \frac{%
q_{0}}{2}\right) ^{2}\int_{t_{0}}^{t}du\int_{t_{0}}^{u}dv\,\sigma
(u)\sigma (v)\sin \omega _{\alpha }(t-u)\sin \omega _{\alpha
}(v-t_{0}) \nonumber
\end{eqnarray}%
%
\begin{equation}
f_{\sigma }=-\frac{c_{\alpha }q_{0}}{m_{\alpha }\omega _{\alpha }}%
\int_{t_{0}}^{t}du\,\sigma (u)\sin \omega _{\alpha }(u-t_{0}),
\end{equation}%
\begin{equation}
g_{\sigma }=-\frac{c_{\alpha }q_{0}}{m_{\alpha }\omega _{\alpha }}%
\int_{t_{0}}^{t}du\,\sigma (u)\sin \omega _{\alpha }(t-u).
\end{equation}%
Combining this result with the its equivalent backward propagator gives
\begin{widetext}
\begin{eqnarray}
\int_{x_{3}t_{0}}^{x't}d[x]\int_{x_{4}t_{0}}^{x't}d[y]\exp\left[\frac{i}{\hbar}
(S_{\alpha I}[\sigma ,x]-S_{\alpha I}[\nu ,y]+S_{\alpha B}[x]-
S_{\alpha B}[y])\right] & = & \nonumber
\\
\frac{m_{\alpha}\omega_{\alpha}}{2\pi\hbar\sin\omega_{\alpha
}(t-t_{0})}
\exp\left(\frac{im_{\alpha}\omega_{\alpha}}{2\hbar\sin\omega_{\alpha
} (t-t_{0})}C\right) & &
\end{eqnarray}
\end{widetext}where
\begin{widetext}
\begin{eqnarray}
  C & = & (x_{3}^{2}-x_{4}^{2})\cos \omega_{\alpha }(t-t_{0}) -
  2x'(x_{3}-x_{4}) \nonumber
 \\
  &  &  - x'[f_{\sigma}(t-t_{0})-f_{\nu}(t-t_{0})] +
 x_{3}g_{\sigma}(t-t_{0})-x_{4}g_{\nu}(t-t_{0})  \nonumber
\\
& & -
\frac{c_{\alpha}^{2}q_{0}^{2}}{m_{\alpha}^{2}\omega_{\alpha}^{2}}
 \int_{t_{0}}^{t}du\int_{t_{0}}^{u}dv \nonumber
\\
& & \times [\sigma (u)\sigma (v)-\nu (u)\nu(v)]
 \sin\omega_{\alpha}(t-u)\sin\omega_{\alpha}(v-t_{0})
\end{eqnarray}
\end{widetext} and the functions $f_{\nu }$ , $g_{\nu }$ are $f_{\sigma }$ ,
$g_{\sigma }$ but with $\nu (u)$ replacing $\sigma (u)$. All of the above is
still just for one oscillator hence the use of the $\alpha $ index. We now
do the integral over $x^{\prime }$ in equation (~\ref{eq:if}), which yields
a delta function, so the path integral part of the influence functional
becomes
\begin{widetext}
 \begin{eqnarray}
\int dx'\int_{x_{3}t_{0}}^{x't}d[x]\int_{x_{4}t_{0}}^{x't}d[y]\exp
\frac{i}{\hbar}(S_{\alpha I}[\sigma ,x]-S_{\alpha I}[\nu
,y]+S_{\alpha B}[x]- S_{\alpha B}[y]) & = &
\nonumber\\
X\,\delta\! \left[ \frac{X}{2}\{2(x_{3}- x_{4}) - (f_{\sigma} -
f_{\nu})\}\right]\exp \left( \frac{iX}{2}D\right) & &
\label{eq:if2},
\end{eqnarray}
\end{widetext}where
\begin{equation}
X=\frac{m_{\alpha }\omega _{\alpha }}{\hbar \sin \omega _{\alpha }(t-t_{0})}
\end{equation}%
and
\begin{widetext}
\begin{eqnarray}
  D & = & (x_{3}^{2}-x_{4}^{2})\cos \omega_{\alpha }(t-t_{0})
 + x_{3}g_{\sigma}(t-t_{0})-x_{4}g_{\nu}(t-t_{0})  \nonumber
\\
& & -
\frac{c_{\alpha}^{2}q_{0}^{2}}{m_{\alpha}^{2}\omega_{\alpha}^{2}}
 \int_{t_{0}}^{t}du\int_{t_{0}}^{u}dv \nonumber
\\
& & \times [\sigma (u)\sigma (v)-\nu (u)\nu(v)]
 \sin\omega_{\alpha}(t-u)\sin\omega_{\alpha}(v-t_{0}) .\label{eq:if3}
\end{eqnarray}
\end{widetext}We note that the above expressions depend on the sign of the
coupling constant $c_{\alpha }$ via the $f_{\sigma ,\nu }$ and $g_{\sigma
,\nu }$ terms, but not the initial value of $\sigma $ (because (\ref{eq:if2}%
) is a product of two propagators). The initial value we chose for $\sigma $%
, $(=+q_{0}/2)$ did, however, appear in the thermalized density matrix (~\ref%
{eq:therm}) for the bath.

We recall from (~\ref{eq:if}) and (~\ref{eq:therm}) the full influence
functional including all the oscillators. The path integrals factor
similarly, so $\int_{{\mbox{{\boldmath$x$}}}_{3}t_{0}}^{{\mbox{{%
\boldmath$x'$}}}t_{0}}d[x]$ is a product of single particle path integrals
over all paths that the individual $x_{\alpha }$ can follow consistent with
the set $\{x_{\alpha }\}$ being ${\mbox{{\boldmath$x'$}}}$ at time t, and ${%
\mbox{{\boldmath$x$}}}_{3}$ at $t_{0}$. If we can \textquotedblleft fold" in
the product of the traces of single-particle propagators given by (\ref%
{eq:if2}) and ~(\ref{eq:if3}) we can then multiply by $\rho ({%
\mbox{{\boldmath$x$}}}_{3}{\mbox{{\boldmath$x$}}}_{4}t_{0})$ and do the
integrals over ${\mbox{{\boldmath$x$}}}_{3}$ and ${\mbox{{\boldmath$x$}}}%
_{4} $ to obtain the influence functional.

\subsection{Evaluation of the influence functional in equilibrium fixed spin
case}

\label{EFSIF}

We want to evaluate a non-standard influence functional, corresponding to
the equilibrium fixed spin initial condition. We do so by replacing it by
the influence functional for a system with the spin and bath uncoupled at $%
t_{0}$ multiplied by a phase due to the fixed spin boundary condition. To
derive this result we now go to sum and difference coordinates $z_{1} =
\frac{x_{3} + x_{4}}{2}$ and $z_{2} = x_{3} - x_{4} $.

If we define $F_{\alpha }[\sigma ,\nu ]$, the single oscillator influence
functional, by $F[\sigma ,\nu ]=\prod_{\alpha }F_{\alpha }[\sigma ,\nu ],$
then using (\ref{eq:if}), and $H_{\alpha 0}$ of the form given in (\ref%
{eq:freeH}), we have
\begin{widetext}
\begin{eqnarray}
 F_{\alpha}[\sigma ,\nu]  & = &\int dz_{1} dz_{2}\,\delta\!\left(z_{1}-
\frac{1}{2}(f_{\sigma} -f_{\nu})\right)<z_{1} + \frac{z_{2}}{2}|
\frac{e^{-\beta H_{\alpha 0}}}{Z_{0}}|z_{1} -\frac{ z_{2}}{2}>
\nonumber
 \\
& & \times\exp\left(
\frac{im_{\alpha}\omega_{\alpha}}{2\hbar\sin\omega_{\alpha
}(t-t_{0})}M\right)
\end{eqnarray}
\end{widetext}and
\begin{eqnarray}
M &=&z_{1}(f_{\sigma }-f_{\nu })\cos \omega _{\alpha
}(t-t_{0})+z_{1}(g_{\sigma }-g_{\nu })+\frac{z_{2}}{2}(g_{\sigma }+g_{\nu })
\nonumber \\
&&-\frac{c_{\alpha }^{2}q_{0}^{2}}{m_{\alpha }^{2}\omega _{\alpha }^{2}}%
\int_{t_{0}}^{t}du\int_{t_{0}}^{u}dv  \nonumber \\
&&\times \lbrack \sigma (u)\sigma (v)-\nu (u)\nu (v)]\sin \omega _{\alpha
}(t-u)\sin \omega _{\alpha }(v-t_{0})
\end{eqnarray}%
where the delta function has been used to replace $x_{3}^{2}-x_{4}^{2}$ by $%
z_{1}(f_{\sigma }-f_{\nu })$. We now consider
\begin{widetext}
\begin{equation}
<z_{1} + \frac{z_{2}}{2}|\frac{\exp{-\beta H_{\alpha
0}}}{Z_{0}}|z_{1} - \frac{ z_{2}}{2}> =
\frac{\rho_{osc}}{Z_{0}}(z_{1} + z_{2}/2 + A ,z_{1} - z_{2}/2 +A )
\end{equation}
\end{widetext}where
\begin{equation}
A=\frac{q_{0}c_{\alpha }}{2m_{\alpha }\omega _{\alpha }^{2}}=\frac{%
q_{0}c_{\alpha }\sigma (t_{0})}{2m_{\alpha }\omega _{\alpha }^{2}}
\end{equation}%
the amount by which the coordinate $x_{\alpha }$ has to be displaced to go
from the free single oscillator density matrix $\rho _{osc}$ to the fixed
spin case. In fact $\rho _{osc}$ is just $\exp \left( -\beta H_{\alpha
}\right) $ with:
\begin{equation}
H_{\alpha }=\frac{p_{\alpha }^{2}}{2m_{\alpha }}+\frac{1}{2}m_{\alpha
}\omega _{\alpha }^{2}x_{\alpha }^{2}.
\end{equation}

If we now write $z_{1} + A = z_{1}^{^{\prime}}$ we get
\begin{equation}
F_{\alpha}[\sigma ,\nu] = F_{\alpha A=0}[\sigma ,\nu]\exp \left( P \right)
\end{equation}
with
\begin{widetext}
\begin{equation}
 P =
\frac
{im_{\alpha}\omega_{\alpha}}{2\hbar\sin\omega_{\alpha}(t-t_{0})}
(-A)\left[(f_{\sigma}-f_{\nu})\cos\omega_{\alpha} (t-t_{0}) +
(g_{\sigma} - g_{\nu})\right],
\end{equation}
\end{widetext}
and $F_{\alpha A=0}$ the conventional influence function~(\ref{eq:waxinf}).

We also have that 
\begin{eqnarray}
& & (f_{\sigma}-f_{\nu})\cos\omega_{\alpha} (t-t_{0}) + (g_{\sigma} -
g_{\nu})  \nonumber \\
& = &
-\frac{c_{\alpha}q_{0}}{m_{\alpha}\omega_{\alpha}}\int_{t_{0}}^{t}du
[\sigma (u) - \nu (u)]\sin\omega_{\alpha}
(t-t_{0})\cos\omega_{\alpha} (u-t_{0}) \nonumber
\end{eqnarray}
so the full influence functional is
\begin{widetext}
\begin{eqnarray}
F[\sigma ,\nu] & = &\left(\prod_{\alpha}F_{\alpha A=0}\right) \exp
\sum_{\alpha}\frac{ic_{\alpha}q_{0}}{2\hbar}A
\int_{t_{0}}^{t}du[\sigma (u) - \nu (u)]\cos\omega_{\alpha}
(u-t_{0})\nonumber
\\
& = & F_{A=0} \exp {\frac{iq_{0}^{2}}{4\hbar}\sigma
(t_{0})\int_{t_{0}}^{t}du [\sigma (u) - \nu (u)]2Q_{1} (u -
t_{0})}\label{eq:fullinf}
\end{eqnarray}
\end{widetext}
where
\begin{equation}
F_{A=0} = \prod_{\alpha}F_{\alpha A=0},
\end{equation}
\begin{eqnarray}
Q_{1}(u) & = & \sum_{\alpha} \frac{c_{\alpha}^{2}}{2m_{\alpha}\omega_{%
\alpha}^{2}}\cos \omega_{\alpha}u  \nonumber  \label{eq:Q1} \\
& = & \int_{0}^{\infty}\frac{d\omega}{\pi}\frac{J(\omega)}{\omega}\cos\omega
u .  \label{eq:q1}
\end{eqnarray}%
and
\begin{equation}
\frac{J(\omega)}{\omega} = \frac{\pi}{2}\sum_{\alpha} \frac{c_{\alpha}^{2}}{%
m_{\alpha}\omega_{\alpha}^{2}}\delta (\omega - \omega_{\alpha})
\end{equation}
is the bath spectral function describing the oscillator heat bath. Equation~(%
\ref{eq:fullinf}) should be compared with the expression arrived at by
Leggett \textit{et al}\cite{Leggett1987} as their equations (B.1.9) and
(B.1.10). All we need now is the standard result
\begin{widetext}
\begin{eqnarray}
F_{A=0}[\sigma , \nu] & = &\exp
-\frac{iq_{0}^{2}}{4\hbar}\int_{t_{0}}^{t}du
\int_{t_{0}}^{u}dv[\sigma (u) - \nu (u)][\sigma (v) + \nu
(v)]Q_{1}^{'}(u-v) \nonumber
\\
& \times & \exp
-\frac{q_{0}^{2}}{4\hbar}\int_{t_{0}}^{t}du\int_{t_{0}}^{u}dv
[\sigma (u) - \nu (u) ][ \sigma (v) - \nu (v) ]Q_{2}(u -
v)\label{eq:waxinf}
\end{eqnarray}
\end{widetext}
where $\tau=\beta \hbar$. We also have the conventionally defined
correlation function $Q_2$ for the fluctuating force (see also Appendix~\ref%
{FlucForce}) and the retarded resistance function $Q_1^{^{\prime}}$ (see
\cite{TakagiBook}) given by
\begin{equation}
Q_{2}(u) = \int_{0}^{\infty}\frac{d\omega}{\pi}J(\omega) \coth\frac{%
\omega\tau}{2}\cos \omega u ,  \label{eq:Q2}
\end{equation}
and
\begin{eqnarray}
Q_{1}^{^{\prime}} (u) & = & \frac{d}{du} Q_{1} (u)  \nonumber \\
& = & -\int_{0}^{\infty}\frac{d\omega}{\pi}\frac{J(\omega)}{\omega}\omega
\sin \omega u
\end{eqnarray}
respectively. The propagator for the reduced spin density matrix in (~\ref%
{eq:reddm2}) becomes
\begin{widetext}
\begin{eqnarray}
K(\sigma_{1}\sigma_{2}t|\sigma_{3}\sigma_{4}t_{0}) & = &
\int_{\sigma_{3}t_{0}}^{\sigma_{1} t}d[\sigma]
\int_{\sigma_{4}t_{0}}^{\sigma_{2}t}d[\nu]A_{n}[\sigma]A_{m}^{*}[\nu]F[\sigma,
\nu]\nonumber
\\
& = & \int_{\sigma_{3}t_{0}}^{\sigma_{1} t}d[\sigma]
\int_{\sigma_{4}t_{0}}^{\sigma_{2}t}d[\nu]A_{n}[\sigma]A_{m}^{*}[\nu]F_{A=0}[\sigma,\nu]
\nonumber
\\
& & \times \exp {\frac{iq_{0}^{2}}{2\hbar}\sigma (t_{0})
\int_{t_{0}}^{t}duQ_{1}(u - t_{0})[\sigma (u) - \nu (u)
]}\label{eq:newprop} .
\end{eqnarray}
\end{widetext}
We see that the equilibrium fixed spin initial condition has altered only
the imaginary (dissipative) first factor in equation 7 of \cite{WaxJPC1985},
Waxman's expression for the influence functional. The second (real) factor
due to fluctuations is unchanged.

\subsection{Simplification of the fixed spin influence functional}

\label{EFSIF2}

We can simplify the first part of the influence functional considerably by
using the extra factor introduced by the spin boundary condition. From~(\ref%
{eq:waxinf}) and (\ref{eq:fullinf}) we have:
\begin{widetext}
\begin{eqnarray}
F[\sigma ,\nu] & = & \exp \{\frac{i}{\hbar}\sigma
(t_{0})(\frac{q_{0}}{2})^{2} \int_{t_{0}}^{t}du[\sigma (u) - \nu
(u)]2Q_{1} (u - t_{0})\}\nonumber
\\
& & \times
\exp\{-\frac{iq_{0}^{2}}{4\hbar}\int_{t_{0}}^{t}du\int_{t_{0}}^{u}dv
[\sigma (u) - \nu (u)][\sigma (v) + \nu (v)]Q_{1}^{'} (u - v
)\}\nonumber
\\
& & \times \exp
-\frac{q_{0}^{2}}{4\hbar}\int_{t_{0}}^{t}du\int_{t_{0}}^{u}dv
[\sigma (u) - \nu (u) ][ \sigma (v) - \nu (v) ]Q_{2}(u -
v)\label{eq:fixinf}.
\end{eqnarray}
\end{widetext}We note that
\begin{widetext}
\begin{eqnarray}
& & \int_{t_{0}}^{u}dv[\sigma (v) + \nu (v)]Q_{1}^{\prime} (u - v)
- 2Q_{1} (u -t_{0})\sigma (t_{0})\nonumber
\\
& = & -[\sigma (u) + \nu (u)]Q_{1} (0) +
\int_{t_{0}}^{u}dv\frac{d}{dv} [\sigma (v) + \nu (v) ]Q_{1} (v -
u)
\end{eqnarray}
\end{widetext}since $\sigma (t_{0})=\nu (t_{0})=1$ . Now
\begin{equation}
\int_{t_{0}}^{t}du[\sigma (u)-\nu (u)][\sigma (u)+\nu (u)]Q_{1}(0)=0
\end{equation}%
as $\sigma ^{2}(u)-\nu ^{2}(u)\equiv 0$, for all $u$ . Therefore the
influence function reduces to
\begin{equation}
F[\sigma ,\nu ]=\exp \left( \int_{t_{0}}^{t}du\int_{t_{0}}^{u}dvf(u,v)\right)
\label{eq:Fuv}
\end{equation}%
with
\begin{eqnarray}
f(u,v) &=&-\frac{iq_{0}^{2}}{4\hbar }[\sigma (u)-\nu (u)]\frac{d}{dv}[\sigma
(v)+\nu (v)]Q_{1}(v-u)  \nonumber \\
&&-\frac{q_{0}^{2}}{4\hbar }[\sigma (u)-\nu (u)][\sigma (v)-\nu
(v)]Q_{2}(u-v)  \label{eq:fuv}
\end{eqnarray}%
which may be compared to equation 7 of \cite{WaxJPC1985}.

We note (see also the appendix \ref{FlucForce}) that the second factor of (%
\ref{eq:fuv}) can be viewed as the effect of a classical fluctuating force.
We will specialise later to the ohmic \cite{Leggett1987} case when
\begin{equation}
J(\omega )\approx \eta \omega e^{-\omega /\omega _{c}}  \label{eq:OhmicJ}
\end{equation}%
for which $Q_{1}(u)$ will be well approximated by a delta function $\eta
\delta (u)$.

We can now follow Waxman \cite{WaxJPC1985} by differentiating the above
propagator $K$ to obtain a differential equation for the time evolution of $%
K $ differing from his equation (15) only in the replacement of $\epsilon$
by $\epsilon(t)$ and by the replacement of his term in $Q_1$ by one arising
from the fixed spin initial condition:
\begin{widetext}
 \begin{eqnarray}
 i\hbar\frac{\partial K}{\partial t} (\sigma_{1}\sigma_{2} t|\sigma_{3}
\sigma_{4} t_{0}) & = &
-\frac{\hbar\Delta}{2}[K(-\sigma_{1}\sigma_{2}
t|\sigma_{3}\sigma_{4} t_{0}) - K(\sigma_{1}-\sigma_{2}
t|\sigma_{3} \sigma_{4} t_{0})]\nonumber
\\
& & +\frac{\hbar\epsilon (t)}{2} (\sigma_{1} - \sigma_{2})
K(\sigma_{1} \sigma_{2} t|\sigma_{3}\sigma_{4} t_{0})\nonumber
\\
& & +\left(\frac{q_{0}}{2}\right)^{2}(\sigma_{1} -
\sigma_{2})\int_{t_{0}}^{t}dv Q_{1}(v-t)\frac{d}{dv}<\sigma (v) +
\nu (v)>\nonumber
\\
& & -i\left(\frac{q_{0}}{2}\right)^{2}(\sigma_{1} -
\sigma_{2})\int_{t_{0}}^{t} dvQ_{2}(v-t)<\sigma (v) - \nu
(v)>\label{eq:Kdif}  .
\end{eqnarray}
\end{widetext}
where
\begin{widetext}
\begin{equation}
<\sigma (v) \pm \nu (v)> =
\int_{\sigma_{3}t_{0}}^{\sigma_{1}t}d[\sigma]
\int_{\sigma_{4}t_{0}}^{\sigma_{2}t}d[\nu]A_{n}[\sigma]A_{m}^{*}[\nu]
F[\sigma ,\nu][\sigma (v) \pm \nu (v)]
\end{equation}
\end{widetext}
For the driven spin-boson system with factorizing initial density matrix and
equilibrium fixed spin initial condition the above integro-differential
equation~(\ref{eq:Kdif}) is exact, and the first main result of this Paper.

\section{Bloch-Redfield equations for the weak damping case}

\label{BlochRedfield}

We now go on in this section (~\ref{subsec:weakdamping}) to demonstrate that
one may obtain a simpler set of equations-of Bloch-Redfield form-for the
spin vector $\mathbf{a}$ by making a weak damping approximation, replacing
terms under the retarded integrals of the propagator equation ~(\ref{eq:Kdif}%
) by their equivalents for an uncoupled spin $\mathbf{a^{(0)}}$. We then (~%
\ref{subsec:dissterms}) identify the dissipation term $\mathbf{{%
\mbox{{\boldmath$\chi$}}}}$ arising from the imaginary factor in the
influence functional. In addition we identify(~\ref{subsec:flucterms}) the
terms due to the fluctuating force allowing us to complete the
Bloch-Redfield equations, and compare them with those of Hartmann \emph{et
al.} \cite{Hartmann2000}(~\ref{subsec:hartmann}).

\subsection{Weak damping approximation}

\label{subsec:weakdamping}

We will now approximate the terms $<\sigma (v) \pm \nu (v)>$ in (\ref%
{eq:Kdif}) by their value for an undamped system i.e. $<\sigma (v) \pm \nu
(v)> \approx <\sigma (v) \pm \nu (v)>_{0} $ This is in general a weak
damping approximation. In the ohmic case it implies $\alpha \ll 1 $ where $%
\alpha $ is the dimensionless friction constant for the problem, $\alpha =
\eta q_{0}^{2} / 2\pi\hbar$ and $\eta$ is as defined in (\ref{eq:OhmicJ}).
We note that linearity implies $<\sigma (v) \pm \nu (v)>_{(0)} = <\sigma
(v)>_{(0)} \pm <\nu (v) >_{(0)}.$ Then $<\sigma (v)>_{0} =
<\sigma_{1}|U^{(0)}(t,v)\sigma_{z}U^{(0)} (v,t_{0})|
\sigma_{3}><\sigma_{2}|U^{(0)}(t,t_{0})|\sigma_{4}>^{*}$ (see also equations
(20) and (21) of \cite{WaxJPC1985}). Here $U^{(0)}$ is the time evolution
operator for a spin uncoupled to the bath:
\begin{equation}
U^{(0)}(t_{1},t_{2}) = \mathcal{T} \exp\left(-\frac{i}{\hbar}%
\int_{t_{1}}^{t_{2}}H_{S} (t^{\prime})dt^{\prime}\right)
\end{equation}
where $\mathcal{T}$ denotes the time-ordering operator.

Using the definition of the inverse of $U^{(0)}$ we find
\begin{widetext}
\begin{equation}
 <\sigma (v)>_{0} = <\sigma_{1}|U^{(0)}(t,v)\sigma_{z}U^{(0)-1}(t,v)U^{(0)}
(t,v)U^{(0)}(v,t_{0})|\sigma_{3}><\sigma_{2}|U^{(0)}(t,t_{0})|\sigma_{4}>^{*}
\end{equation}
\end{widetext} but $U^{(0)}(t,v)U^{(0)}(v,t_{0})=U^{(0)}(t,t_{0})$ so $%
<\sigma (v)>_{0}=<\sigma _{1}|U^{(0)}(t,v)\sigma
_{z}U^{(0)-1}(t,v)U^{(0)}(t,t_{0})|\sigma _{3}><\sigma
_{2}||U^{(0)}(t,t_{0})|\sigma _{4}>^{\ast }.$ Inserting a complete set of
states of $\sigma _{z}$:
\begin{equation}
1=\sum_{\sigma }|\sigma ><\sigma |=|+\sigma _{1}><+\sigma _{1}|+|-\sigma
_{1}><-\sigma _{1}|
\end{equation}%
we have:
\begin{widetext}\bea
<\sigma (v)
>_{0} & =  &
<\sigma_{1}|U^{(0)}(t,v)\sigma_{z}U^{(0)-1}(t,v)|\sigma_{1}>K^{(0)}(
\sigma_{1} \sigma_{2} t|\sigma_{3}\sigma_{4} t_{0})\nonumber
\\
& & +
<\sigma_{1}|U^{(0)}(t,v)\sigma_{z}U^{(0)-1}(t,v)|-\sigma_{1}>K^{(0)}(-
\sigma_{1} \sigma_{2} t|\sigma_{3}\sigma_{4}
t_{0})\label{eq:zeroth}   \eea \end{widetext} where the superscript on $%
K^{(0)}$ indicates that it propagates the uncoupled spin density matrix. If
\begin{equation}
K(\sigma _{1}\sigma _{2}t|\sigma _{3}\sigma _{4}t_{0})=\int_{\sigma
_{3}t_{0}}^{\sigma _{1}t}d[\sigma ]\int_{\sigma _{4}t_{0}}^{\sigma
_{2}t}d[\nu ]A_{n}[\sigma ]A_{m}^{\ast }[\nu ]F[\sigma ,\nu ]
\end{equation}%
then
\begin{equation}
K^{(0)}(\sigma _{1}\sigma _{2}t|\sigma _{3}\sigma _{4}t_{0})=\int_{\sigma
_{3}t_{0}}^{\sigma _{1}t}d[\sigma ]\int_{\sigma _{4}t_{0}}^{\sigma
_{2}t}d[\nu ]A_{n}[\sigma ]A_{m}^{\ast }[\nu ]
\end{equation}%
i.e. the case $F[\sigma ,\nu ]=1$ (no environment). This is important
because $K$ propagates $\tilde{\rho}(\sigma _{3}\sigma _{4}t_{0})$ to $%
\tilde{\rho}(\sigma _{1}\sigma _{2}t)$ (with $t>t_{0}$) while
$K^{(0)}$ will propagate the same $\tilde{\rho}(\sigma _{3}\sigma
_{4}t_{0})$ to a
different final density matrix which we have called $\tilde{\rho}%
^{(0)}(\sigma _{1}\sigma _{2}t)$. Neglect of this point would lead
to the appearance of terms $O(\alpha ^{2})$ in the final equation
of motion. We note that the propagator $J$ in equation (22) of
\cite{WaxJPC1985} should, in the notation of the present paper,
have been written as $J^{(0)}$.

We now want to evaluate the terms
\begin{equation}
<\sigma _{1}|U^{(0)}(t,v)\sigma _{z}U^{(0)-1}(t,v)|\sigma _{1}>.
\label{eq:bracket}
\end{equation}%
To do this consider the density operator $\rho ^{(0)}(t)$ which is the
solution of
\begin{equation}
i\hbar \frac{\partial }{\partial t}\rho ^{(0)}(t)=[H_{S},\rho ^{(0)}]
\end{equation}%
Then $\rho ^{(0)}(t)=U^{(0)}(t,t_{0})\rho ^{(0)}(t_{0})U^{(0)-1}(t,t_{0})$
allowing us to form the identity $2\rho ^{(0)}(t)-1=U^{(0)}(t,t_{0})[2\rho
^{(0)}(t_{0})-1]U^{(0)-1}(t,t_{0}).$ Defining a polarization vector ${%
\mbox{{\boldmath$a$}}}^{(0)}(t)$ by $\rho ^{(0)}(t)=\frac{1}{2}(1+{%
\mbox{{\boldmath$a$}}}^{(0)}(t)\cdot {\mbox{{\boldmath$\sigma$}}})$ we can
write $2\rho -1$ as
\begin{equation}
{\mbox{{\boldmath$a$}}}^{(0)}(t)\cdot {\mbox{{\boldmath$\sigma$}}}%
=U^{(0)}(t,t_{0})[{\mbox{{\boldmath$a$}}}^{(0)}(t_{0})\cdot {%
\mbox{{\boldmath$\sigma$}}}]U^{(0)-1}(t,t_{0}).  \label{eq:at}
\end{equation}%
If ${\mbox{{\boldmath$a$}}}^{(0)}(t_{0})\cdot {\mbox{{\boldmath$\sigma$}}}%
=\sigma _{z}$ which is the case we need in order to evaluate (\ref%
{eq:bracket}), we can then (c.f. \cite{WaxJPC1985}) define an auxiliary
polarization vector ${\mbox{{\boldmath$f$}}}(t,t_{0})$ by ${%
\mbox{{\boldmath$f$}}}(t,t_{0})={\mbox{{\boldmath$a$}}}^{(0)}(t)$. The
dependence on $t_{0}$ enters via (\ref{eq:at}). ${\mbox{{\boldmath$f$}}}%
(t,t_{0})$ corresponds to the polarization vector for an isolated spin
evolved from time $t_{0}$ to time $t$ subject to the initial condition that
its polarization vector was $(0,0,1)$ at time $t_{0}$. We will later use the
equivalent notation ${\mbox{{\boldmath$a$}}}^{(0)}(t,t_{0})$ for this
function, where the above initial condition is implied.

Having established the meaning of the terms $<\sigma_{1}|U(t,v) \sigma_{z}
U^{-1}(t,v)|\sigma_{1}>$ we are in a position to substitute the expressions
for them into the equation of motion for $K(\sigma_{1} \sigma_{2}
t|\sigma_{3}\sigma_{4} t_{0})$. We first note that:
\begin{equation}
<\sigma_{1}|U(t,v){\mbox{{\boldmath$a$}}}^{(0)}(v)\cdot{\mbox{{\boldmath$%
\sigma$}}}U^{-1}(t,v)| \sigma_{1}> = \sigma_{1}f_{3}(t,v)
\end{equation}
and similarly
\begin{equation}
<\sigma_{1}|U(t,v){\mbox{{\boldmath$a$}}}^{(0)}(v)\cdot{\mbox{{\boldmath$%
\sigma$}}}U^{-1}(t,v)| -\sigma_{1}> = f_{1}(t,v) - i\sigma_{1}f_{2}(t,v)
\end{equation}
Hence equation (\ref{eq:zeroth}) becomes:
\begin{widetext}\be
 <\sigma (v)>_{0} = (f_{1}(t,v) - if_{2}(t,v)\sigma_{1})K^{(0)}(-\sigma_{1}
\sigma_{2} t|\sigma_{3}\sigma_{4} t_{0}) + \sigma_{1}f_{3}(t,v)
K^{(0)}(\sigma_{1}\sigma_{2} t|\sigma_{3}\sigma_{4}
t_{0})\label{eq:sig0}  \ee \end{widetext}
This step is crucial because it allows the use of the spin polarization
vectors for an uncoupled spin.

Substitution of our expression for $\partial_{t} K(\sigma_{1}\sigma_{2} t|
\sigma_{3}\sigma_{4} t_{0})$ into the equation for $\partial_{t}\tilde{\rho}
(\sigma_{1}\sigma_{2} t)$ will now give a soluble expression for the reduced
density matrix $\tilde{\rho}(\sigma_{1}\sigma_{2} t)$. Using equations (\ref%
{eq:reddm2}), (\ref{eq:sig0}) and (\ref{eq:Kdif}) we have:
\begin{widetext}
\bea i\hbar\frac{\partial}{\partial
t}\tilde{\rho}(\sigma_{1}\sigma_{2} t) & = &
 i\hbar\frac{\partial}{\partial t}\sum_{\sigma_{3}\sigma_{4}}K(\sigma_{1}
\sigma_{2} t|\sigma_{3}\sigma_{4}
t_{0})\tilde{\rho}(\sigma_{3}\sigma_{4} t_{0}) \nonumber
\\
& = & i\hbar\sum_{\sigma_{3}\sigma_{4}}\partial_t K (
\sigma_{1}\sigma_{2} t|\sigma_{3}\sigma_{4} t_{0}) \tilde{\rho}
(\sigma_{3}\sigma_{4} t_{0})\nonumber
\\
& = & -\frac{\Delta}{2}\hbar[\tilde{\rho}(-\sigma_{1}\sigma_{2} t)
- \tilde{\rho}(\sigma_{1} -\sigma_{2} t)]
 +\epsilon (t)\frac{\hbar}{2}(\sigma_{1} - \sigma_{2})\tilde{\rho}
(\sigma_{1}\sigma_{2} t)\nonumber
\\
& & + \left(\frac{q_{0}}{2}\right)^{2}(\sigma_{1} - \sigma_{2})
\nonumber
\\
& & \times [\int_{t_{0}}^{t}dvQ_{1}(v-t)
\frac{d}{dv}\{f_{1}(t,v)\tilde{\rho}^{(0)}(-\sigma_{1}\sigma_{2}
t) + f_{1}(t,v) \tilde{\rho}^{(0)}(\sigma_{1}-\sigma_{2}
t)\}\nonumber
\\
& & + \int_{t_{0}}^{t}dvQ_{1}(v-t)
\frac{d}{dv}\{-if_{2}(t,v)(\sigma_{1}\tilde{\rho}^{(0)}(-\sigma_{1}\sigma_{2}
t) - \sigma_{2}\tilde{\rho}^{(0)}(\sigma_{1}-\sigma_{2}
t))\}\nonumber
\\
& & +(\sigma_{1} + \sigma_{2}) \int_{t_{0}}^{t}dvQ_{1}(v-t)
\frac{d}{dv}f_{3}(t,v)\tilde{\rho}^{(0)}(\sigma_{1}\sigma_{2}
t)]\nonumber
\\
& & -i\int_{t_{0}}^{t}dvQ_{2}(t-v)
\{f_{1}(t,v)\tilde{\rho}^{(0)}(-\sigma_{1}\sigma_{2} t) -
f_{1}(t,v) \tilde{\rho}^{(0)} (\sigma_{1}-\sigma_{2} t)\}\nonumber
\\
& & - \int_{t_{0}}^{t}dvQ_{2}(t-v)
\{f_{2}(t,v)\sigma_{1}\tilde{\rho}^{(0)}(-\sigma_{1}\sigma_{2} t)
-
 f_{2}(t,v)\sigma_{2}\tilde{\rho}^{(0)}(\sigma_{1}-\sigma_{2} t)\}\nonumber
\\
& & -i(\sigma_{1} - \sigma_{2}) \int_{t_{0}}^{t}dvQ_{2}(t-v)
f_{3}(t,v)\sigma_{1}\tilde{\rho}^{(0)}(\sigma_{1}\sigma_{2} t) ]
 +O(\alpha^{2})
\label{eq:rhodif}. \eea
\end{widetext}

\subsection{Identification of spin precession terms}

\label{sec:precession}

We note first that, after defining the reduced density operator $\tilde{\rho}%
(t)$ and its corresponding polarization vector ${\mbox{{\boldmath$a$}}}(t)$
through
\begin{equation}
\tilde{\rho}(\sigma_{1}\sigma_{2} t) = <\sigma_{1}|\tilde{\rho}|\sigma_{2}>
= <\sigma_{1}|\frac{1}{2} (1+{\mbox{{\boldmath$a$}}}(t)\cdot{%
\mbox{{\boldmath$\sigma$}}})|\sigma_{2}>.
\end{equation}
the terms in $\Delta$ and $\epsilon(t)$ on the right of (\ref{eq:rhodif})
will give the familiar spin precession term
\begin{equation}
\frac{\partial {\mbox{{\boldmath$a$}}}}{\partial t} = {\mbox{{\boldmath$b$}}}%
(t)\wedge{\mbox{{\boldmath$a$}}} + \ldots
\end{equation}
We have also defined ${\mbox{{\boldmath$b$}}}(t) = (-\Delta,0,\epsilon(t)) $
so that
\begin{equation}
H_{S} = \frac{\hbar}{2}{\mbox{{\boldmath$b$}}}(t)\cdot{\mbox{{\boldmath$%
\sigma$}}}.
\end{equation}

\subsection{Identification of dissipation terms}

\label{subsec:dissterms}

We now consider the next three groups of terms, i.e. those in $Q_{1}$, in
equation~(\ref{eq:rhodif}). These are due to dissipation and one may call
them the systematic terms by analogy with the Fokker-Planck equation. The
last is zero as $(\sigma _{1}-\sigma _{2})(\sigma _{1}+\sigma _{2})=\sigma
_{1}^{2}-\sigma _{2}^{2}\equiv 0$ for all $\sigma _{1},\sigma _{2}$
(remember $\sigma _{i}=\pm 1$ for $i=1,2$). The first can be found by
forming commutators:
\begin{widetext}
\bea
   (\sigma_{1}-\sigma_{2})[\tilde{\rho}^{(0)}(-\sigma_{1}\sigma_{2} t) +
 \tilde{\rho}^{(0)}
(\sigma_{1}-\sigma_{2} t)] & = &
<\sigma_{1}|[\sigma_{z},\sigma_{x}\tilde{\rho}^{(0)}(t)]|\sigma_{2}>
\nonumber \\
& & +
<\sigma_{1}|[\sigma_{z},\tilde{\rho}^{(0)}(t)\sigma_{x}]|\sigma_{2}>
. \eea
\end{widetext} Now $\tilde{\rho}^{(0)}(t)=\frac{1}{2}(1+{\mbox{{%
\boldmath$a$}}}^{(0)}(t)\cdot {\mbox{{\boldmath$\sigma$}}})$ so the above
expression becomes $<\sigma _{1}|2i\sigma _{y}|\sigma _{2}>.$ Similarly the
second term gives $-i(\sigma _{1}-\sigma _{2})[\sigma _{1}\tilde{\rho}%
(-\sigma _{1}\sigma _{2}t)-\sigma _{2}\tilde{\rho}(\sigma _{1}-\sigma
_{2}t)]=<\sigma _{1}|-2i\sigma _{x}|\sigma _{2}>$. Thus if
\[
\frac{d}{dv}f_{1}(t,v)=f_{1}^{^{\prime }}(t,v)
\]%
and similarly for $f_{2}^{\prime }$ and $f_{3}^{\prime }$ we have
\begin{widetext}
\bea i\hbar\frac{\partial}{\partial
t}\frac{1}{2}\mvec{a}\cdot\mvec{\sigma} & = &
 \ldots \nonumber
\\
& &
+\left(\frac{q_{0}}{2}\right)^{2}\int_{t_{0}}^{t}dv\,Q_{1}(t-v)2i\{f_{1}^{'}
(t,v)\sigma_{y} - f_{2}^{'}(t,v)\sigma_{x}\}\nonumber
\\
& & + \ldots \eea
\end{widetext} where we have used the evenness of $Q_{1}(u)\equiv Q_{1}(-u)$%
. If $e_{y}$ is a unit vector in the y direction then ${\mbox{{\boldmath$%
\sigma$}}}\cdot {\mbox{{\boldmath$e$}}}_{y}=\sigma _{y}$ etc., and defining
a vector ${\mbox{{\boldmath$f$}}}^{^{\prime }}(t,v)$ by
\begin{equation}
{\mbox{{\boldmath$f$}}}^{^{\prime }}(t,v)=(f_{1}^{^{\prime
}},f_{2}^{^{\prime }},f_{3}^{^{\prime }})=\frac{d}{dv}{\mbox{{\boldmath$f$}}}%
(t,v)
\end{equation}%
we then have
\begin{equation}
\frac{\partial {\mbox{{\boldmath$a$}}}}{\partial t}={\mbox{{\boldmath$b$}}}%
(t)\wedge {\mbox{{\boldmath$a$}}}+{\mbox{{\boldmath$\chi$}}}+\ldots
\label{eq:adot}
\end{equation}

with
\begin{equation}
{\mbox{{\boldmath$\chi$}}} = \frac{4}{\hbar}\left(\frac{q_{0}}{2}%
\right)^{2}\int_{t_{0}}^{t}dv\,Q_{1} (t-v){\mbox{{\boldmath$e$}}}_{z}\wedge{%
\mbox{{\boldmath$f$}}}^{^{\prime}}(t,v).  \label{eq:chidef}
\end{equation}

This is completes the derivation of the dissipative part of the
Bloch-Redfield equations for this initial condition.

\subsection{Identification of fluctuating force terms}

\label{subsec:flucterms}

We finally have the group of terms in $Q_2$ from equation (~\ref{eq:rhodif}%
). First
\begin{eqnarray}
& & (\sigma_{1} - \sigma_{2})(\tilde{\rho}^{(0)}(-\sigma_{1}\sigma_{2} t) -
\tilde{\rho}^{(0)} (\sigma_{1}-\sigma_{2} t))  \nonumber \\
& = & \frac{1}{2}<\sigma_{1}|a^{(0)}_{y}(t)[\sigma_{z},[\sigma_{x},%
\sigma_{y}]]| \sigma_{2} >  \nonumber \\
& & +\frac{1}{2}<\sigma_{1}|a^{(0)}_{z}(t)[\sigma_{z},[\sigma_{x},%
\sigma_{z}]]| \sigma_{2}>  \nonumber \\
& = & 0 +<\sigma_{1}|a^{(0)}_{z}(t)(2i)^{2}\sigma_{x}|\sigma_{2}>
\end{eqnarray}
and similarly
\begin{widetext}
\be
 -i(\sigma_{1} - \sigma_{2})[\sigma_{1}\tilde{\rho}^{(0)}
(-\sigma_{1}\sigma_{2} t) +
\sigma_{2}\tilde{\rho}^{(0)}(\sigma_{1}-\sigma_{2} t)]
 =  <\sigma_{1}|-2\sigma_{y}a^{(0)}_{z}(t)|\sigma_{2}>
\ee
\end{widetext}
while
\begin{equation}
(\sigma_{1} - \sigma_{2})^{2}\tilde{\rho}(\sigma_{1}\sigma_{2} t) =
<\sigma_{1}|(2\sigma_{x}a_{x}^{(0)}(t) +
2\sigma_{y}a_{y}^{(0)}(t)|\sigma_{2}>.
\end{equation}
If we substitute these values into equation (\ref{eq:rhodif}) and use the
auxiliary vector ${\mbox{{\boldmath$f$}}}$ we find that the equation of
motion for ${\mbox{{\boldmath$a$}}}(t)$ is:
\begin{eqnarray}
\frac{\partial {\mbox{{\boldmath$a$}}}}{\partial t} & = & {%
\mbox{{\boldmath$b$}}}\wedge{\mbox{{\boldmath$a$}}}  \nonumber \\
& & +\frac{4}{\hbar}\left(\frac{q_{0}}{2}\right)^{2}\int_{t_{0}}^{t}dv%
\,Q_{1} (t-v){\mbox{{\boldmath$e$}}}_{z}\wedge{\mbox{{\boldmath$f$}}}%
^{^{\prime}}(t,v)  \nonumber \\
& & +{\mbox{{\boldmath$e$}}}_{z}\wedge\frac{4}{\hbar}\left(\frac{q_{0}}{2}
\right)^{2}\int_{t_{0}}^{t}dv \,Q_{2}(v - t){\mbox{{\boldmath$f$}}}%
(t,v)\wedge{\mbox{{\boldmath$a$}}}^{(0)}(t)  \nonumber \\
& = &{\mbox{{\boldmath$b$}}}\wedge{\mbox{{\boldmath$a$}}} + {%
\mbox{{\boldmath$\chi$}}} - {\mbox{{\boldmath$e$}}}_{z}\wedge({%
\mbox{{\boldmath$\psi$}}}(t) \wedge{\mbox{{\boldmath$a$}}}^{(0)})
+O(\alpha^{2})  \label{eq:waxy}
\end{eqnarray}
where we have defined the retarded integral
\begin{equation}
\psi_{j} = -\frac{4}{\hbar}\left(\frac{q_{0}}{2}\right)^{2}%
\int^{t}_{t_{0}}dv\, Q_{2}(t - v)f_{j}(t,v)  \label{eq:psidef}
\end{equation}

Equation~(\ref{eq:waxy}) has been derived by omitting terms of $O(\alpha^{2})
$ and higher. While this procedure seems reasonable it leads, at long times,
to unreasonable results in, for example, problems where ${%
\mbox{{\boldmath$b$}}}$ is independent of time. In these cases the ${%
\mbox{{\boldmath$a$}}}^{(0)}$ term continues to oscillate indefinitely, and
thus the system will not tend to a time independent state of equilibrium, as
we believe it should. The simple replacement of ${\mbox{{\boldmath$a$}}}%
^{(0)}$ by ${\mbox{{\boldmath$a$}}}$ appears to cure this problem; for a
static ${\mbox{{\boldmath$b$}}}$ the system does equilibrate \cite%
{WaxJPC1985}. We view this as a selective resummation (to infinite order in $%
\alpha$) that is sufficient to guarantee more correct long time behaviour,
which we make due to its physical reasonableness.

We thus replace ${\mbox{{\boldmath$a$}}}^{0}(t)$ in equation~(\ref{eq:waxy})
by ${\mbox{{\boldmath$a$}}}(t)$ to give:

\begin{equation}
\frac{\partial {\mbox{{\boldmath$a$}}}}{\partial t} = {\mbox{{\boldmath$b$}}}%
(t)\wedge{\mbox{{\boldmath$a$}}} + {\mbox{{\boldmath$\chi$}}} - \hat{{%
\mbox{{\boldmath$e$}}}}_{z}\wedge({\mbox{{\boldmath$\psi$}}}(t) \wedge{%
\mbox{{\boldmath$a$}}})  \label{eq:watkins}
\end{equation}

Equation~(\ref{eq:watkins}) and its derivation by path integral, with the
auxiliary definitions of ${\mbox{{\boldmath$\chi$}}}$ (~\ref{eq:chidef}) and
$\psi$ (~\ref{eq:psidef}) forms the second main result of our work.

\subsection{Comparison with Bloch-Redfield equations of Hartmann \textit{et
al.}}

\label{subsec:hartmann}

Hartmann \textit{et al.}\cite{Hartmann2000} quote a set of Bloch-Redfield
equations obtained from projection operator methods as their equation (4):

\begin{eqnarray}
\left(
\begin{array}{l}
{\dot \sigma_x} \\
{\dot \sigma_y} \\
{\dot \sigma_z}%
\end{array}
\right) & = & \left(
\begin{array}{lll}
0 & \epsilon & 0 \\
-\epsilon & 0 & \Delta \\
0 & -\Delta & 0%
\end{array}
\right) \left(
\begin{array}{l}
{\ \sigma_x} \\
{\ \sigma_y} \\
{\ \sigma_z}%
\end{array}
\right)  \nonumber \\
& + & \left(
\begin{array}{lll}
-\Gamma_{xx} & 0 & -\Gamma_{xz} \\
0 & -\Gamma_{yy} & -\Gamma_{yz} \\
0 & 0 & 0%
\end{array}
\right) \left(
\begin{array}{l}
{\ \sigma_x} \\
{\ \sigma_y} \\
{\ \sigma_z}%
\end{array}
\right)  \nonumber \\
& + & \left(
\begin{array}{l}
{A_x} \\
{A_y} \\
{0}%
\end{array}
\right)
\end{eqnarray}

where $\sigma _{x,y,z}$ are spin expectation values i.e. $a_{x,y,z}$ in our
notation. To establish the correspondence with our work note first that they
have a complex correlation function:
\begin{eqnarray}
M &=&\frac{1}{\pi }\int_{0}^{\infty }d\omega J(\omega )[\cosh (\hbar \omega
/2k_{B}T-i\omega t)/\sinh (\hbar \omega /2k_{B}T)]  \nonumber \\
&=&M^{^{\prime }}+iM^{^{\prime \prime }}
\end{eqnarray}%
which is equivalent to $Q_{2}+iQ_{1}^{\prime }$ in our notation. They also
define transition amplitudes for an uncoupled spin between right and right,
and left and right, pairs of states as $U_{RR}=<R|U(t,t^{^{\prime }})|R>$
and $U_{RL}=<R|U(t,t^{^{\prime }}|L>$ respectively, and an auxiliary
function $F(t)=2\int_{0}^{t}dt^{^{\prime }}M^{^{\prime \prime
}}(t-t^{^{\prime }})U_{RR}U_{RL}$ so that $A_{x}= Im F(t)$ and $%
A_{y}= Re F(t)$. Finally they have the damping tensors $\Gamma
_{ij}$ defined by $\Gamma _{ij}=\int_{0}^{t}dt^{^{\prime
}}M^{^{\prime
}}(t-t^{^{\prime }})b_{ij}(t,t^{^{\prime }})$. If we follow Hartmann \textit{%
et al.} by defining $b_{yz}=-2 Im (U_{RR}U_{RL})$ and
$b_{xz}=-2Re(U_{RR}U_{RL})$, it becomes clear that their $b_{yz}$
corresponds to our $a_{y}^{0}$ and $b_{xz}$ to our $a_{x}^{0}$.

We thus find that their $\Gamma_{xx}=\Gamma_{yy} \equiv-\psi_z $, and also $%
\Gamma_{xz} \equiv \psi_x$, $\Gamma_{yz} \equiv \psi_y$. This gives an
identical structure to ours for the fluctuating force term, while their
dissipation term ${\mbox{{\boldmath$A$}}}=A_x \hat{e}_x + A_y \hat{e}_y
\left( \equiv {\mbox{{\boldmath$\chi$}}}\right)$ is different, being
\begin{equation}
A_x \equiv - \int_{t_0}^t dt^{^{\prime}} Q_1^{^{\prime}} (t-t^{^{\prime}})
a_y^{0}
\end{equation}
\begin{equation}
A_y \equiv \int_{t_0}^t dt^{^{\prime}} Q_1^{^{\prime}} (t-t^{^{\prime}})
a_x^{0}
\end{equation}

However the difference would disappear if we were to revert to a factorizing
initial condition without fixed spin. The other point to stress is that both
our derivation and theirs arrive at a fluctuating force term with $a^{0}$,
the substitution of $a$ at that point has to be an additional assumption.


\section{Bloch-Redfield Equations in the ohmic and zero temperature cases}

\label{OhmicBR}

We now use the result of Waxman \citep{WaxJPC1985} and Zhang %
\citep{Zhang1988} that in the ohmic case the full expression for the time
dependent ${\mbox{{\boldmath$\chi$}}}$ vector (~\ref{eq:chidef}) greatly
simplifies to the time independent $\chi \hat{e}_{x}$. In the critical
damping case studied by Shytov \citep{Shytov2000} where $J(\omega )\sim
\omega ^{-1}$ both $x$ and $y$ terms would be present in ${%
\mbox{{\boldmath$\chi$}}}$.

\subsection{Simplification of the dissipation term for ohmic spectral
function}

In the ohmic case, equation (\ref{eq:Q1}) becomes $Q_{1}(u)\approx \eta
\delta (u)$ so
\begin{equation}
{\mbox{{\boldmath$\chi$}}}=\frac{\eta q_{0}^{2}}{2\hbar }\hat{e}_{z}\wedge {%
\mbox{{\boldmath$f$}}}^{\prime }(t,t)  \label{eq:chiohm}
\end{equation}%
where the extra factor of 1/2 comes from the use of the delta function at an
endpoint of the integral. Now
\begin{equation}
{\mbox{{\boldmath$f$}}}^{\prime }(t,t)\equiv \lim_{t^{\prime }\rightarrow t}%
\frac{d}{dt^{\prime }}{\mbox{{\boldmath$f$}}}(t^{\prime },t)={%
\mbox{{\boldmath$b$}}}\wedge {\mbox{{\boldmath$f$}}}(t,t)
\end{equation}%
where the last step follows from the fact that
\begin{equation}
\frac{d}{dt^{\prime }}{\mbox{{\boldmath$f$}}}(t^{\prime },t)={%
\mbox{{\boldmath$b$}}}\wedge {\mbox{{\boldmath$f$}}}(t^{\prime },t).
\label{eq:fmotion}
\end{equation}%
As
\begin{equation}
\lim_{t^{\prime }\rightarrow t}{\mbox{{\boldmath$f$}}}(t^{\prime },t)={%
\mbox{{\boldmath$f$}}}(t,t)=(0,0,1)
\end{equation}%
we have
\begin{equation}
{\mbox{{\boldmath$b$}}}\wedge {\mbox{{\boldmath$f$}}}(t,t)=(-\Delta
,0,\epsilon (t))\wedge (0,0,1)=\Delta {\mbox{{\boldmath$e_{y}$}}}
\end{equation}%
so
\begin{equation}
{\mbox{{\boldmath$e_{z}$}}}\wedge ({\mbox{{\boldmath$b$}}}\wedge {%
\mbox{{\boldmath$f$}}}(t,t))=\Delta {\mbox{{\boldmath$e_{z}$}}}\wedge {%
\mbox{{\boldmath$e_{y}$}}}=-\Delta {\mbox{{\boldmath$e_{x}$}}}.
\end{equation}%
We thus find that the dissipation term (\ref{eq:chiohm}) becomes:
\begin{eqnarray}
{\mbox{{\boldmath$\chi$}}} &=&\frac{\eta q_{0}^{2}}{2\hbar }(-\Delta ){%
\mbox{{\boldmath$e_{x}$}}}  \nonumber \\
&=&-\pi \alpha \Delta {\mbox{{\boldmath$e_{x}$}}}  \nonumber \\
&=&\chi {\mbox{{\boldmath$e_{x}$}}}  \label{eq:chiterm}
\end{eqnarray}

We note that this term acts in the negative ${\mbox{{\boldmath$e_{x}$}}}$
direction. By taking the dot product of ${\mbox{{\boldmath$a$}}}$ with (\ref%
{eq:watkins}) to find $\partial{\mbox{{\boldmath$a$}}}^2/\partial t $ we see
that the dissipation term thus tends to reduce the length of the vector ${%
\mbox{{\boldmath$a$}}}$, in contrast to the magnitude-conserving spin
precession term. This is directly analogous to the way that an electric
field can change a speed while a magnetic field cannot, in the Lorenz force
law.

\subsection{Additional simplification of fluctuation term in zero
temperature limit:}

We have, provided that $\omega_c \tau \gg 1$,
\begin{equation}
Q_2(v)=\frac{\eta}{\pi}[\frac{\omega_c^{-2}-v^2}{(w^{-2}_c+v^2)^2} + \frac{1%
}{v^2}- (\frac{\tau}{\pi}\sinh\frac{\pi v }{\tau})^{-2}]
\end{equation}

This has two components  corresponding to zero and finite
temperatures respectively. We will here specialize to the zero
temperature limit. In this case $\psi _{3}$ becomes $-2\pi \alpha
$ while $\psi _{x}$ and $\psi _{y}$ tend to zero.


\section{Application to Landau-Zener-Stuckelberg problem:}

\label{Plots}

The dissipative LZS problem as usually considered (e.g. \cite{A&R1989})
corresponds to the driven spin-boson Hamiltonian with the specific choice $%
\epsilon (t)=\epsilon t$, although other choices may be of interest (e.g.
\cite{NakamuraRedBook}). We emphasise that in this paper we have kept
general time dependence $\epsilon (t)$ of~(\ref{eq:spinbosonH}) up to now.
The standard LZS choice corresponds to a spin modelling a particle in a
biased double potential well for which an external driving force acts to
reverse the sign of the offset between the wells at $t=0$; $t$ having
increased from a large negative starting value $t=t_{0}$. The quantitities
of interest are usually the long time, e.g. $t\rightarrow \infty $, values
of the probability $P_{LZ}$ that the spin is in the \textquotedblleft down" (%
$-1$) orientation if started \textquotedblleft up" (the conventional
\textquotedblleft tunnelling probability") or, equivalently, the long
positive time expectation values $a_{x},a_{y},a_{z}$ of the spin operator $%
\sigma $.

In absence of coupling to environment the LZS tunnelling probability can be
obtained exactly as (e.g. \cite{NakamuraRedBook,Berry}):

\begin{equation}
P_{LZS}=1-e^{-\pi \Delta^2/2 \epsilon} = 1-e^{-\pi \frac{\Delta}{2} \tau_z}
\end{equation}
where $\tau_z = \Delta/\epsilon$ is the Zener (characteristic tunnelling)
time.

$P_{LZS}$ is $\sim 1-e^{- S/\hbar}$ where the action $S$ is the energy for
one hop times the Zener time, which, to within a numerical factor, can be
derived from an instanton calculation \cite{Shytov2000}. The inverse of the
tunnelling angular frequency expressed in units of $\tau_z$ is $2/(\Delta
\tau_z) = \lambda$, a nonadiabaticity parameter \cite{Berry}, which
increases as the speed of change of the energy levels increases i.e. as the
system becomes more nonadiabatic.

Figures 1 and 2 illustrate the Landau-Zener behaviour (see also \cite{Berry}%
). For large $\lambda $ the value of $S$ tends to zero, as does $P_{LZS}$
and so the probability of not tunnelling tends to $1$ i.e. in the fast
passage limit the system does not tunnel. In the opposite limit of
adiabatically slow change, the particle clings to the instantaneous
eigenstate and must tunnel, so $P_{LZS}$ tends to $1$, i.e. the probability
of not tunnelling tends to $0$.

\subsection{Illustrative numerical solutions of ohmic Bloch-Redfield
equation at zero temperature}

To solve our Bloch-Redfield equations we need to rescale time to the Zener
time $\tau_z$. After doing this the ohmic version becomes:
\begin{equation}
\frac{\partial{\mbox{{\boldmath$a$}}}}{\partial T} = {\mbox{{\boldmath$B$}}}%
\wedge{\mbox{{\boldmath$a$}}}(T) + \frac{2\pi\alpha}{\lambda}e_{x} -
e_{z}\wedge({\mbox{{\boldmath$\psi$}}}(T)\wedge{\mbox{{\boldmath$a$}}}(T))
\label{eq:scaledode}
\end{equation}

We first note that this form shows why the effect of the dissipation term is
negligible for the (fast passage) non-adiabatic limit of large $\lambda$. It
is simply because the dissipation term is inversely proportional to $\lambda$%
. This means that (within the limits of our approximation) the faster a
damped qubit is switched, the less it will be affected.

We show representative solutions of~(\ref{eq:scaledode}) in the absence of
fluctuations as an illustration of the size of the perturbation of the
probability caused by the spin-bath coupling. Figure 3 shows how $%
(1/2)(1+a_z)$ changes in the most adiabatic case studied, $\lambda=1$,
comparing the uncoupled evolution (blue curve) with $\alpha= 0.001$ Figure 4
is for the same case as Figure 3 but shows the 3 components of spin, only
for the damped case. Figures 5 and 6 are analogous but show the fast passage
$\lambda=15$ case. Figures 7 and 8 are also for the fast passage case $%
\lambda=15$ but are for stronger coupling to the bath $\alpha=0.01$. At this
level of coupling the damping of quantum oscillations is very noticeable,
although it does not alter the final tunnelling probability.


\section{Conclusions}

\label{SumUp} This paper has extended the static spin boson study of Waxman
\cite{WaxJPC1985} to the driven spin boson case, obtaining an exact
integro-differential equation for the time evolution of the propagator of
the reduced spin density matrix, the first main result. By specializing to
weak damping we then obtained the next result, a set of Bloch-Redfield
equations for the equilibrium fixed spin initial condition. Finally we
showed that these equations can be used to solve the classic dissipative
Landau-Zener problem and illustrated these solutions for the weak damping
case. The effect of dissipation was seen to be minimised as passage speed
increased, implying that qubits need to be switched as fast as possible.


\begin{acknowledgments}
We are grateful to Sandra Chapman, Patrick Espy, Mark Madsen, George
Rowlands and Andrey Shytov for valuable interactions. NWW thanks Norman
Dombey, Gabriel Barton, Irwin Shapiro, Mervyn Freeman and Mai Mai Lam for
encouraging him to complete this study. NWW acknowledges the support of SERC
and PPARC at the Universities of Sussex and Warwick where portions of this
work were performed, and the recent hospitality of Tom Chang at the Center
for Space Research at MIT while it was being completed.
\end{acknowledgments}

\appendix

\section{Notation}

$\alpha$ dimensionless coupling constant \newline
${\mbox{{\boldmath$a$}}}(t,t_0)$ spin polarisation vector\newline
${\mbox{{\boldmath$a^{(0)}$}}}(t,t_0)$ ($\equiv {\mbox{{\boldmath$f$}}}%
(t,t_0)$)spin polarisation vector in uncoupled case\newline
$A[\sigma]$ Spin amplitude functional for path containing $n$ flips\newline
$A[\sigma,\tilde{f}]$ Spin amplitude in stochastic case\newline
$\beta$ Inverse temperature\newline
${\mbox{{\boldmath$b$}}}$ vector derived from 2-level Hamiltonian \newline
$\{c_{\alpha}\}$ Set of spin-bath coupling constants $c_{\alpha}$\newline
$d[x]$ measure for path integral\newline
$d[\sigma]$ measure for spin path integral \newline
$\Delta$ tunnelling matrix element\newline
$D$ Correlation function in stochastic case \newline
$\epsilon (t)$ bias\newline
$\eta$ ohmic coupling strength \newline
$\tilde{f}$ Fluctuating force \newline
$F[\sigma,\nu]$ Influence functional\newline
$F_{A=0}$ Influence functional in uncoupled case\newline
$\hat{e}_x,\hat{e}_y,\hat{e}_z$ unit vectors \newline
$\hbar$ Planck constant \newline
$H(t)$ Hamiltonian with corresponding action $S[\sigma,x]$\newline
$H_0$ ($\equiv H(t=t_0)$)\newline
$H_B(t)$ Bath Hamiltonian with corresponding action $S_B[x]$\newline
$H_I(t)$ Interaction Hamiltonian with corresponding action $S_I[\sigma,x]$%
\newline
$H_S(t)$ Spin Hamiltonian with corresponding action $S_S[\sigma]$\newline
$J(\omega)$ bath spectral function\newline
$K^{(0)}(t)$ spin propagator for $\tilde{\rho}^{(0)}$\newline
$K(t)$ spin propagator for $\tilde{\rho}$\newline
${\mbox{{\boldmath$\chi$}}}$ dissipation term \newline
$k_B$ Boltzmann constant\newline
$\lambda$ adiabaticity parameter \newline
$m_{\alpha}$ oscillator mass\newline
$\omega_{\alpha}$ oscillator frequency \newline
$\omega_c$ bath cutoff \newline
$\psi_{i}$ $i=x,y,z$ \newline
$P_{LZS}$ Landau-Zener tunnelling probability \newline
$P[\tilde{f}]$ stochastic force probability functional \newline
$q_0$ inter-well spacing \newline
$Q_1$ retardation function \newline
$Q_2$ fluctuating force correlation function \newline
$\rho_{osc}$ Single oscillator density matrix\newline
$\rho$ density matrix \newline
$\tilde{\rho}$ spin density matrix\newline
$\sigma_{x,y,z}$ Pauli spin matrices \newline
$\tau_z$ Zener time \newline
$t$ time\newline
$t_0$ fiducial time \newline
$T$ temperature \newline
$U$ unitary time evolution operator \newline
$\{x_{\alpha}\}$ (or ${\mbox{{\boldmath$x$}}}$) set of harmonic oscillator
positions $x_{\alpha}$ \newline
$Z_0$ partition function \newline
$z_1,z_2$ temporary sum and difference co-ordinates\newline


\section{Identification of Fluctuating Force and Derivation of Stochastic
Equation of Motion for Propagator of driven spin-boson system}

\label{FlucForce}

We now show that the second term in the influence functional is equivalent
to a random fluctuating force leading to an alternative equation of motion
for the propagator of stochastic (quantum Fokker-Planck) type. This result
is still quite general, but we confirm that the known delta-correlated force
would be recovered in the high temperature limit for an ohmic choice of
spectral density.

We first define a probability functional $P[\tilde{f}]$ which gives the
probability that a member of a statistical ensemble experiences a force $%
\tilde{f}$ at a given instant:
\begin{equation}
P[f] = \exp -\frac{1}{2}\int \tilde{f}(u)D^{-1}(u,v) \tilde{f}(v)
\end{equation}
normalised such that
\begin{equation}
\int d[\tilde{f}]P[\tilde{f}] = <1> = 1.
\end{equation}
Then we define an average over the ensemble of forces
\begin{equation}
<e^{-\int \tilde{f}(u)h(u)}>_{\tilde{f}} = e^{\frac {1}{2}\int
h(u)D(u,v)h(v)}
\end{equation}
which implies that the autocorrelation function is given by
\begin{equation}
<\tilde{f}(u)\tilde{f}(v)> = D(u,v).
\end{equation}
If we expand $D(u,v)$ in terms of its eigenvalues $\{\lambda_{n}\}$
\begin{equation}
D(u,v) = \sum_{n} \lambda_{n}\phi_{n}(u)\phi_{n}(v)
\end{equation}
we can define its inverse $D^{-1}(u,v)$ by
\begin{equation}
D^{-1}(u,v) = \sum_{n} \lambda_{n}^{-1}\phi_{n}(u)\phi_{n}(v)
\end{equation}
where we require $\lambda_{n} > 0$ for all $n$. We define the correlation
function $Q_{2}(u,v)$ through
\begin{equation}
D(u,v) = 2\hbar Q_{2}(u-v).
\end{equation}
Hence if we take the average
\begin{eqnarray}
<e^{\frac{iq_{0}}{2\hbar}\int_{t_0}^{t}du[\sigma(u)-\nu(u)]\tilde{f}
(u)}>_{f} & = & \nonumber \\
e^{-\frac{1}{\hbar}(\frac{q_{0}}{2})^{2}\int_{t_0}^{t}du
\int_{t_0}^{t}dv[\sigma(u)
-\sigma(u)][\sigma(v)-\sigma(v)]Q_{2}(u-v)} & &
\end{eqnarray}
we obtain an object which has the the same form as the second, real factor
in the influence functional~(\ref{eq:fuv}). The identification of the
correlation function $Q_{2}(u-v)$ can be made with our previously defined $%
Q_{2}(u-v)$~(\ref{eq:Q2}) provided that the Fourier transform is positive.
We recall that $Q_{2}$ is of the form of a cosine transform so we can see
positivity is satisfied.

We can thus define a new amplitude $A[\sigma, \tilde{f}]$ and an influence
functional $\tilde{F}[\sigma, \nu]$ via
\begin{equation}
A[\sigma, \tilde{f}] = (\frac{i\Delta}{2})^{n}\exp -\frac{i}{\hbar }%
\int_{t_0}^{t}du( \frac{\hbar \epsilon}{2} + \frac{q_{0}\tilde{f}(u)}{2}%
)\sigma(u)
\end{equation}
and
\begin{equation}
\tilde{F}[\sigma, \nu] = \exp -\frac{i}{\hbar}(\frac{q_{0}}{2})^{2}
\int_{t_0}^{t}du\int_{t_0}^{t}dv[\sigma(u)-\sigma(u)][\sigma(v)+\nu(v)]Q^{%
\prime}_{1}(u-v).
\end{equation}
The stochastic analogue of~(\ref{eq:Kdif}) is, before averaging with respect
to $\tilde{f}$:
\begin{widetext}
\begin{eqnarray}
i\hbar\partial_{t}\tilde{K}(\sigma_{1}\sigma_{2}t|\sigma_{3}\sigma_{4})
& = &
-\frac{\hbar\Delta}{2}[\tilde{K}(-\sigma_{1}\sigma_{2}t|\sigma_{3}\sigma_{4}0)
-
\tilde{K}(\sigma_{1}-\sigma_{2}t|\sigma_{3}\sigma_{4}0)]\nonumber \\
&  & + \left(\frac{\hbar\epsilon (t)}{2} +
\frac{q_{0}}{2}\tilde{f}(t)\right)(\sigma_{1}
-\sigma_{2})K(\sigma_{1}\sigma_{2}t|\sigma_{3}\sigma_{4}0)\nonumber \\
&  &
+\left(\frac{q_{0}}{2}\right)^{2}(\sigma_{1}-\sigma_{2})\int_{0}^{t}dvQ'_{1}
(t-v)<\sigma(v)+\nu(v)>, \label{eq:StocKdif}\end{eqnarray}
\end{widetext}
where we denoted the stochastic propagator by $\tilde{K}$. This propagator
equation could now be used with~(\ref{eq:reddm2}) to obtain a quantum
Fokker-Planck equation for the driven spin-boson model, analogously with
\cite{WaxJPC1985,ChangWaxman}.

The replacement of the $Q_2$ term has thus contributed a fluctuating bias to
the existing time dependent bias. In consequence the stochastic analogue of
the Bloch-Redfield equation~(\ref{eq:watkins}), before averaging with
respect to $\tilde{f}$, becomes:
\begin{equation}
\frac{\partial {\mbox{{\boldmath$a$}}}}{\partial t} = {\mbox{{\boldmath$b$}}}%
(t)\wedge{\mbox{{\boldmath$a$}}} + {\mbox{{\boldmath$\chi$}}} + \frac{q_{0}}{%
\hbar}\tilde{f}(t)\hat{e}_{z}\wedge{\mbox{{\boldmath$a$}}}
\label{eq:ohmadot}
\end{equation}

This is a weak damping quantum Langevin equation. Here $<\tilde{f}(u)\tilde{f%
}(v)>_{\tilde{f}} = 2\hbar Q_{2}(u-v)$ and $<{\mbox{{\boldmath$a$}}}>_{%
\tilde{f}}$ is the physical quantity. $\tilde{f}(t)$ has the physically
appealing interpretation of being the fluctuating force on the spin
associated with a given member of the ensemble of systems considered. We
believe that this may be more tractable than the Bloch-Redfield form in
cases when the retarded integrals would otherwise need to be evaluated.

We note that we can recover the known delta-correlated behaviour of $Q_2$
for the high temperature ($\tau \rightarrow 0$), ohmic spectral density (Eq.~%
\ref{eq:OhmicJ}) limit. Using (~\ref{eq:OhmicJ}) we first substitute for $%
J(w)$ in (~\ref{eq:Q2}), and then, provided the bath cutoff frequency $w_c$
is sufficiently large but $\omega \tau=\hbar \omega / k_B T$ still
sufficiently small, we can approximate $\coth x$ by $1/x$. This gives us $%
Q_2 = [ \frac{2 \eta k_{B} T}{\pi \hbar} ] \delta (u)$ and thus
\begin{equation}
< \tilde{f}(t) \tilde{f}(t^{\prime}) > = \frac{4 \eta k_B T}{\pi} \delta (t
- t^{\prime})
\end{equation}

\newpage

\section{Green's function solution of ohmic quantum Langevin equation}

\label{GreenFunc}

We here solve the quantum Langevin equation perturbatively using
the Green function method of Zhang \citep{Zhang1988}. We find the
intriguing result in the ohmic weak damping case that the
fluctuations enter only at second order in $\alpha$, and so give
an expression for the tunnelling probability to first order in
$\alpha$.

We use the methods of equations (46-52) of \citep{Zhang1988}, but unlike
Zhang we do not differentiate the Green's function solution.

We have:

\begin{equation}
\frac{\partial \mathbf{a}}{\partial t} = \mathbf{b}(t) \wedge \mathbf{a} +
\mathbf{\chi} + \frac{q_0}{\hbar} \tilde{f}(t) \hat{e}_z \wedge \mathbf{a}
\end{equation}
where $\tilde{f(t)}$ is the fluctuating force. This allows us to write a
solution of the stochastic equation as
\begin{equation}
\mathbf{a}(t, \tilde{f}) = \mathcal{M}(t,t_0) \mathbf{a}(t_0) +
\int_{t_0}^{t} \mathcal{M}(t,t^{\prime}) \mathbf{\chi} (t^{\prime})
dt^{\prime}
\end{equation}
(c.f. equation (46) of \citep{Zhang1988}) with $\mathcal{M}$ a Green
function for the homogeneous equation

\begin{equation}
\frac{\partial \mathbf{a}}{\partial t} - \mathbf{b} \wedge \mathbf{a} -
\frac{q_0}{\hbar} \tilde{f}(t) \hat{e}_z \wedge \mathbf{a} = 0
\end{equation}

We then take out the evolution operator $u(t,t^{\prime})$ for the
unperturbed spin, and then (equation (49) of \cite{Zhang1988} rearranged)
note that
\begin{equation}
u(t,t^{\prime})\mathcal{N}(t,t^{\prime}) = \mathcal{M}(t,t^{\prime})
\end{equation}
giving
\begin{equation}
\mathbf{a}(t, f) = u(t,t_0)\mathcal{N}(t,t_0)\mathbf{a}(t_0) +
\int_{t_0}^{t} u(t,t^{\prime})\mathcal{N}(t,t^{\prime}) \mathbf{\chi}
(t^{\prime}) dt^{\prime}
\end{equation}

Now, following equation (52) of \citep{Zhang1988}, the formal solution for $%
\mathcal{N}$ is
\begin{equation}
\mathcal{N} (t,t^{\prime}) = 1 + \sum \int_{t^{\prime}}^{t} dt_1 ...
\int_{t^{\prime}}^{t_{n-1}} dt_n \mathcal{F}(t_1,t^{\prime}) \mathcal{F}
(t_n, t^{\prime})
\end{equation}
with
\begin{equation}
\mathcal{F} = \frac{q_0}{\hbar} \tilde{f}(t) e_z \wedge .
\end{equation}
If we now average over the stochastic force, we will have
\begin{widetext}
\begin{equation}
{\bf a}(t) = < {\bf a} (t,\tilde{f}) >_{\tilde{f}} =
u(t,t_0)<{\cal N}(t,t_0)>{\bf a}(t_0) + \int_{t_0}^{t}
u(t,t')<{\cal N}(t,t')> {\bf \chi} (t') dt' \label{eq:GFint}
\end{equation}
\end{widetext}
Now $<\mathcal{N}(t,t_0)>$ is
\begin{widetext}
\begin{equation}
1 + \int_{t_0}^{t} dt_1 <{\cal F}(t_1,t')> +  \int_{t_0}^{t}
\int_{t_0}^{t_1} dt_1 dt_2 <{\cal F}(t_1,t') {\cal F}(t_2,t')>
\end{equation}
\end{widetext} to second order in $\mathcal{F}$ (which we know will be first
order in $\alpha$). We may take $<\mathcal{F}(t_1,t^{\prime})>$ to be zero
for the fluctuating force without serious loss of generality.

We note that
\begin{widetext}
\begin{eqnarray}
<{\cal N}(t,t_0)> {\bf a}(t_0) & = & [1 +
\int_{t_0}^{t} \int_{t_0}^{t_1} dt_1 dt_2 <{\cal F}(t_1,t') {\cal F}(t_2,t')>]{\bf a}(t_0) \\
& = & [ 1 + \frac{q_0^2}{\hbar^2}
\int_{t_0}^{t} \int_{t_0}^{t_1} dt_1 dt_2 < \tilde{f}(t_1) \tilde{f}(t_2) > \hat{e}_z \wedge ( \hat{e}_z \wedge ] {\bf a}(t_0) \nonumber \\
& = & {\bf a}(t_0)
\end{eqnarray}
\end{widetext}because [$\hat{e}_{z}\wedge (\hat{e}_{z}\wedge ]\mathbf{a}%
(t_{0})=0$ for the case we chose, where $a(t_{0})$ was $(0,0,1)$. The terms
under the Green's function integral (i.e. equation(~\ref{eq:GFint}) for $%
\mathcal{N}(t,t^{\prime })\mathbf{\chi }(t^{\prime })$ are all second order
in $\alpha $ except the first one which from above expansion can be seen to
be just $\mathbf{\chi }(t^{\prime })$, so we are left with a very simple
solution

\begin{eqnarray}
\mathbf{a}(t) & = & u(t,t_0) \mathbf{a}(t_0) + \int_{t_0}^{t}
u(t,t^{\prime}) \mathbf{\chi} (t^{\prime}) dt^{\prime} \\
& = & \mathbf{a^{0}}(t, t_0) + \chi \int_{t_0}^{t} \mathbf{a^{1(0)}}%
(t,t^{\prime}) dt^{\prime}
\end{eqnarray}
where $\mathbf{a^{1(0)}}(t,t^{\prime})$ is just the undamped solution $%
\mathbf{a^{0}} $ at time $t$, but started at $(1, 0, 0)$ at $t^{\prime}$. In
the ohmic case after rescaling times by $\tau_z$ we have $\chi = 2 \pi
\alpha / \lambda$ (c.f. \ref{eq:scaledode} ) so we have a linear dependence
on $\alpha$ for a given $\lambda$. Checking this prediction with numerical
solutions for $\alpha = 0.005$ agrees very well with the above result. We
note this is a consequence of the spin-up initial condition, as others might
give a first order contribution from the fluctuating force. It is however
independent of the spectral form of the environment once that initial
condition has been specified.

\newpage


\begin{figure*}[tbp]
\includegraphics{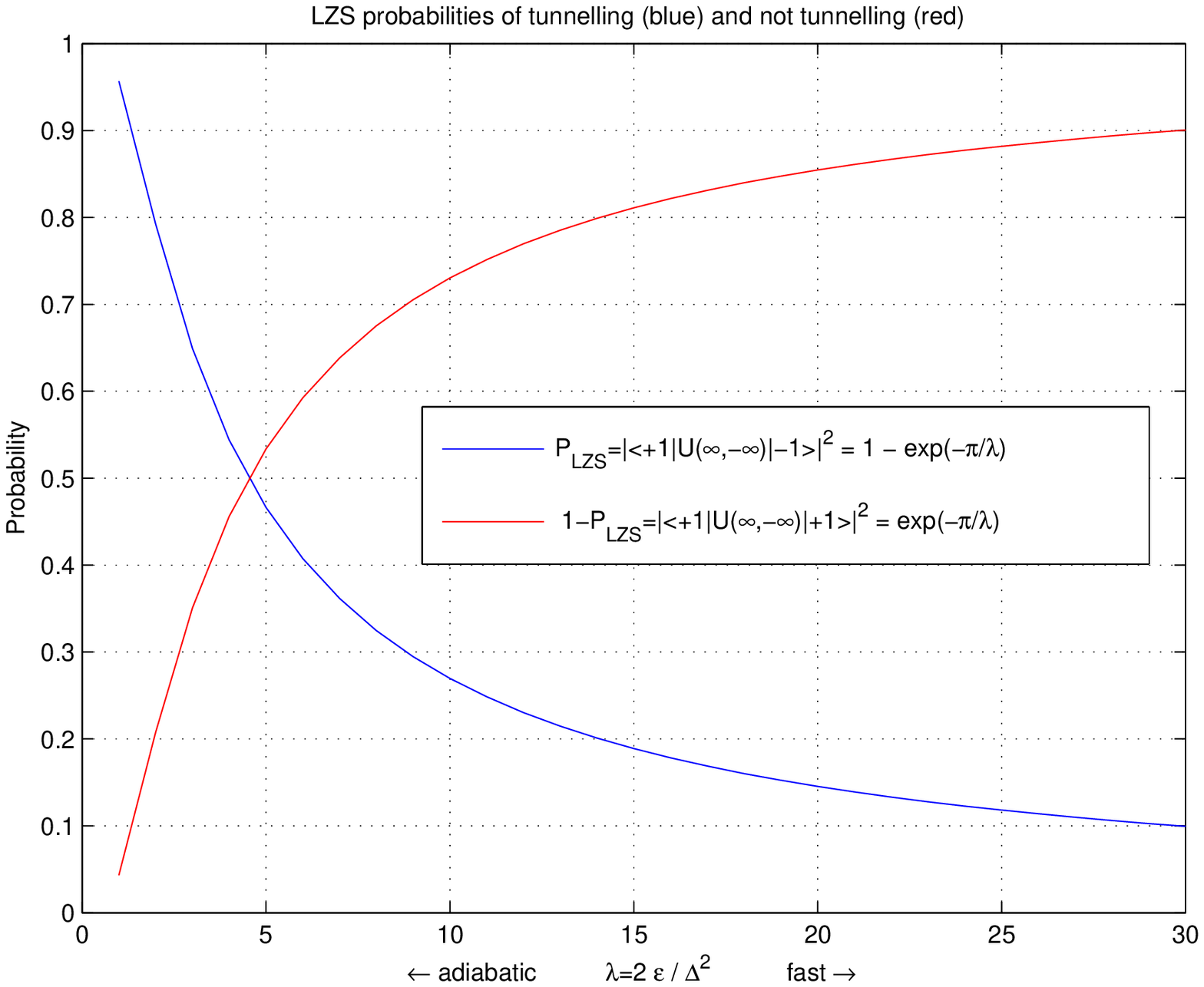}
\caption{Dependence of the Landau-Zener-Stuckelberg tunnelling
probability $P_{LZS}$ and
the probability of not tunnelling $1-P_{LZS}$ on the adiabaticity parameter $%
\protect\lambda $ in the absence of spin-bath coupling.}
\label{fig:wide}
\end{figure*}

\begin{figure*}[tbp]
\includegraphics{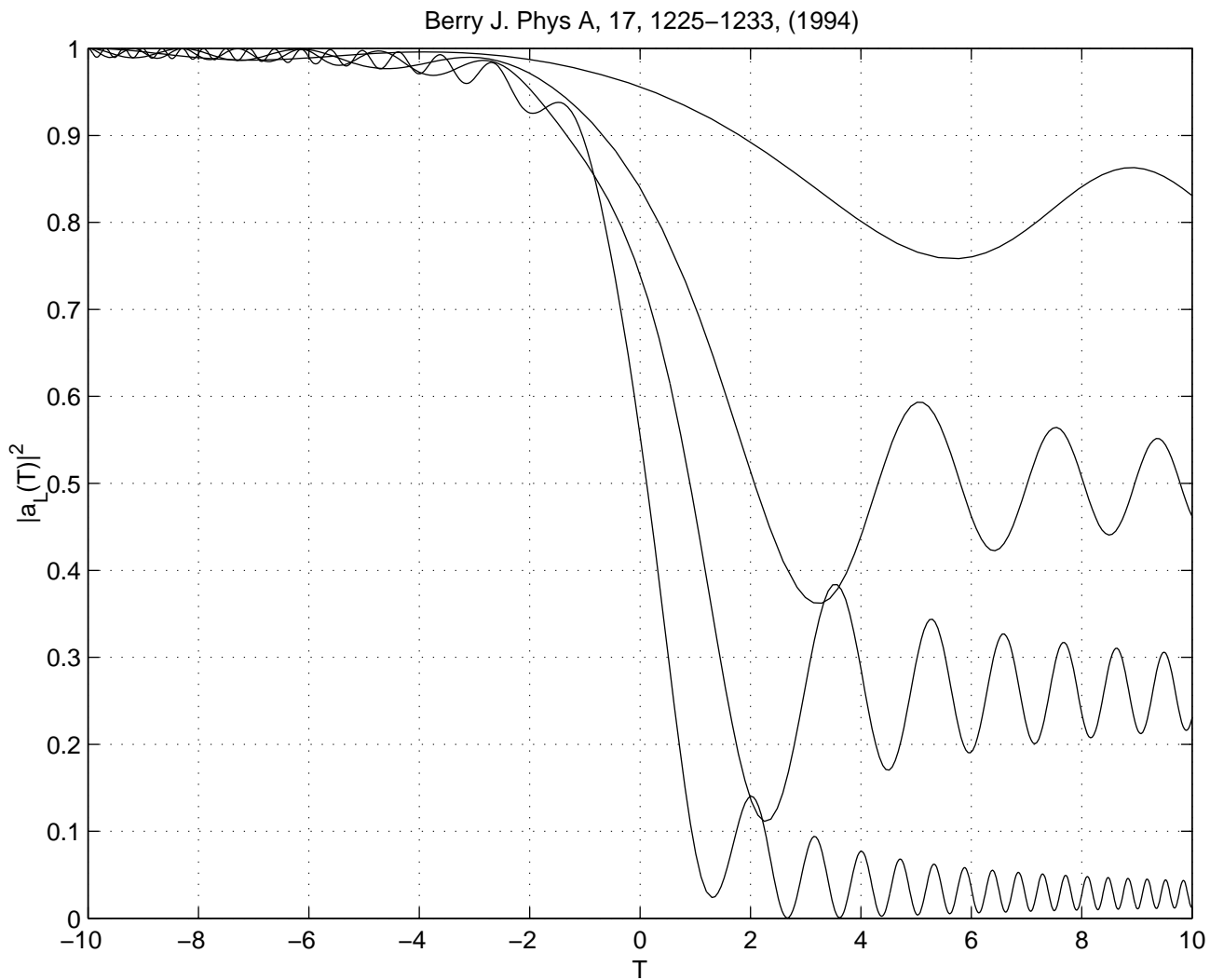}
\caption{Evolution from $t=-10 \protect\tau_z$ of $1-P_{LZS}$ to
$t=10 \protect\tau_z$ for spins started with $\mathbf{a}=(0,0,1)$
in the absence of spin-bath coupling (after Berry
\protect\cite{Berry}). $\protect\lambda$ increases as the curves
go up the plot, taking values of $1, 2.5, 5, 15$ }
\label{fig:wide}
\end{figure*}

\begin{figure*}[tbp]
\includegraphics{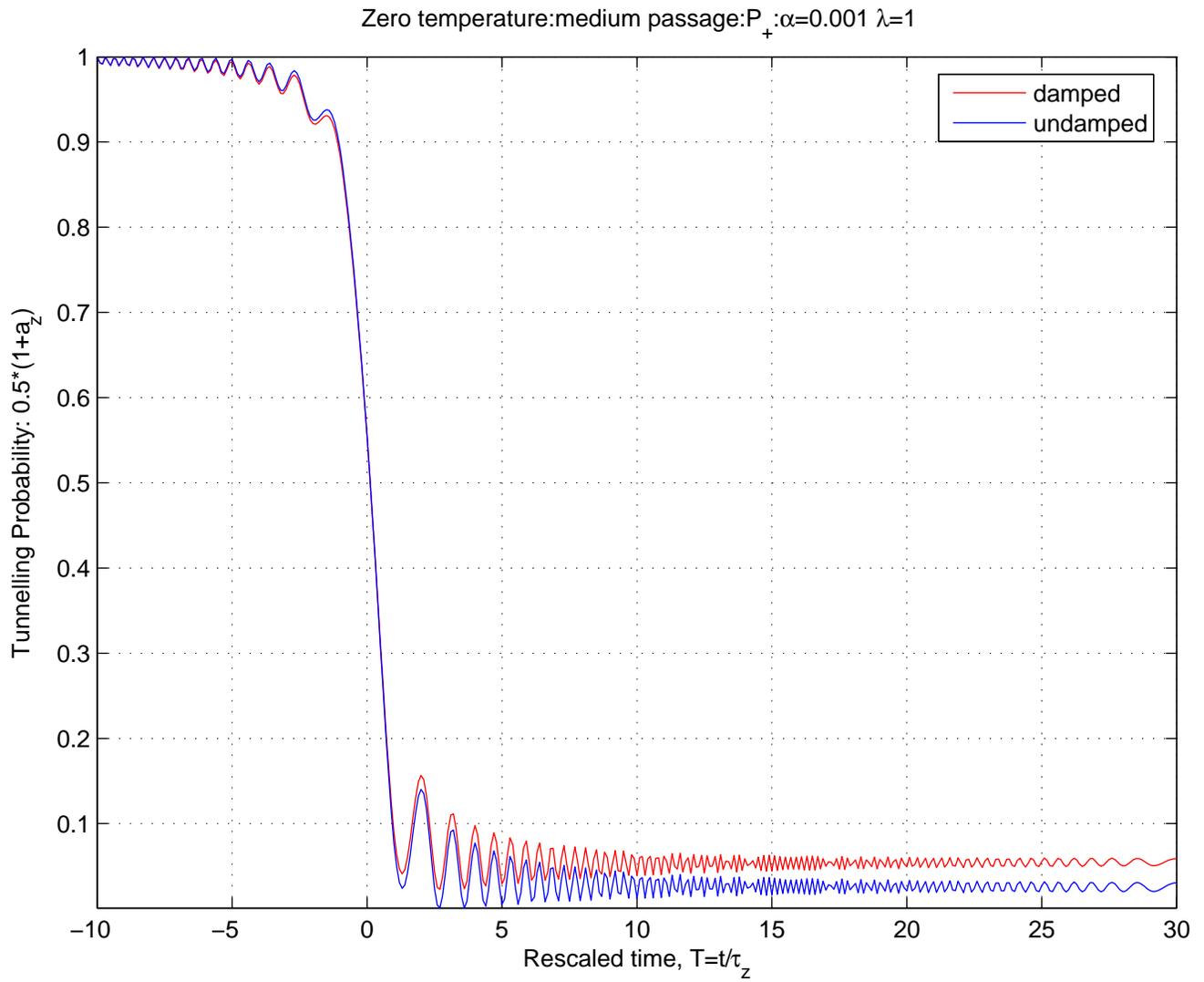}
\caption{The evolution of $(1/2)(1+a_z)$ with time in medium passage case, $%
\protect\lambda=1$, comparing the uncoupled evolution (blue curve) with $%
\protect\alpha= 0.001$ (red curve).}
\label{fig:wide}
\end{figure*}

\begin{figure*}[tbp]
\includegraphics{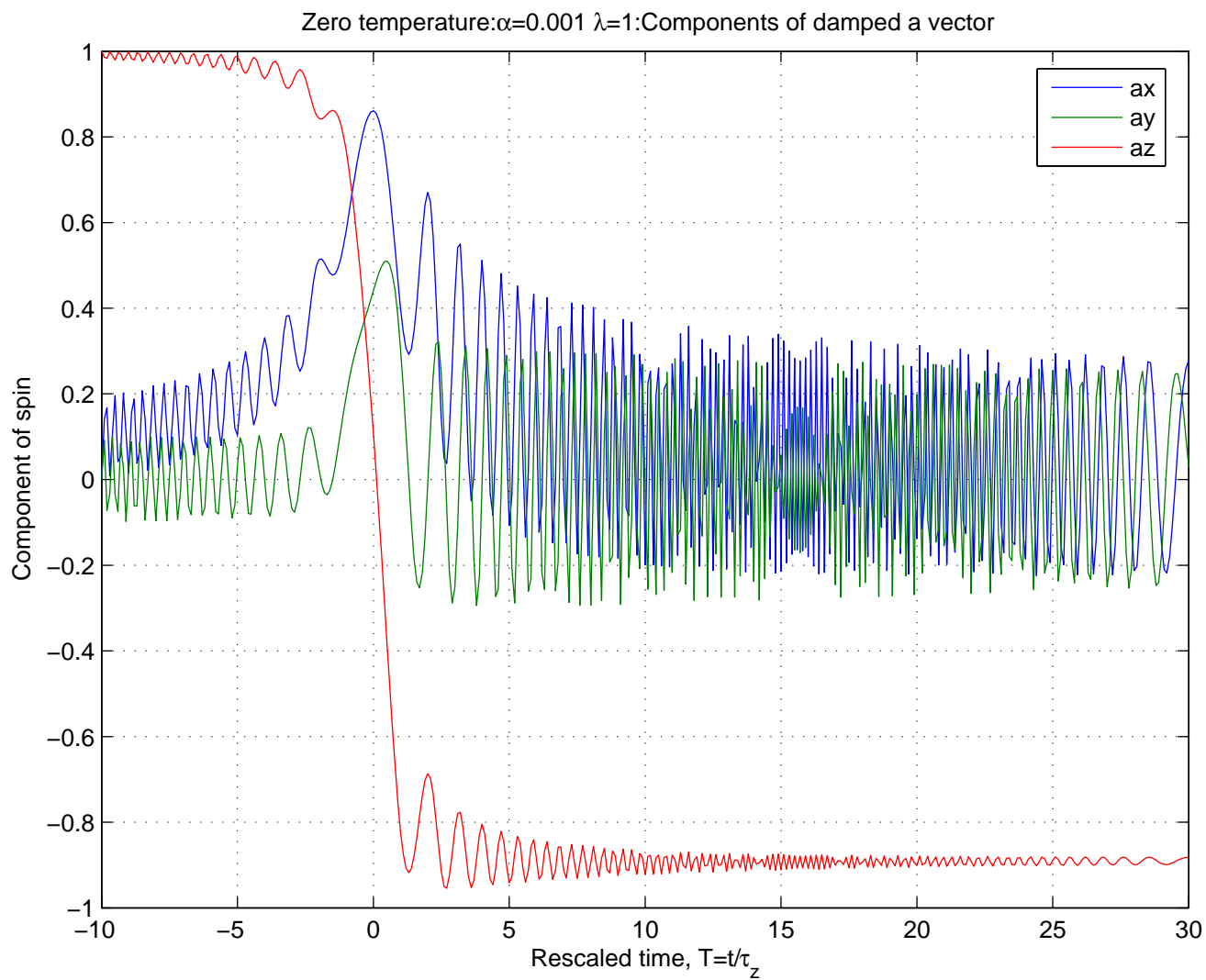}
\caption{Three components of spin for the same case as Figure 3 showing only
the damped cases. }
\label{fig:wide}
\end{figure*}

\begin{figure*}[tbp]
\includegraphics{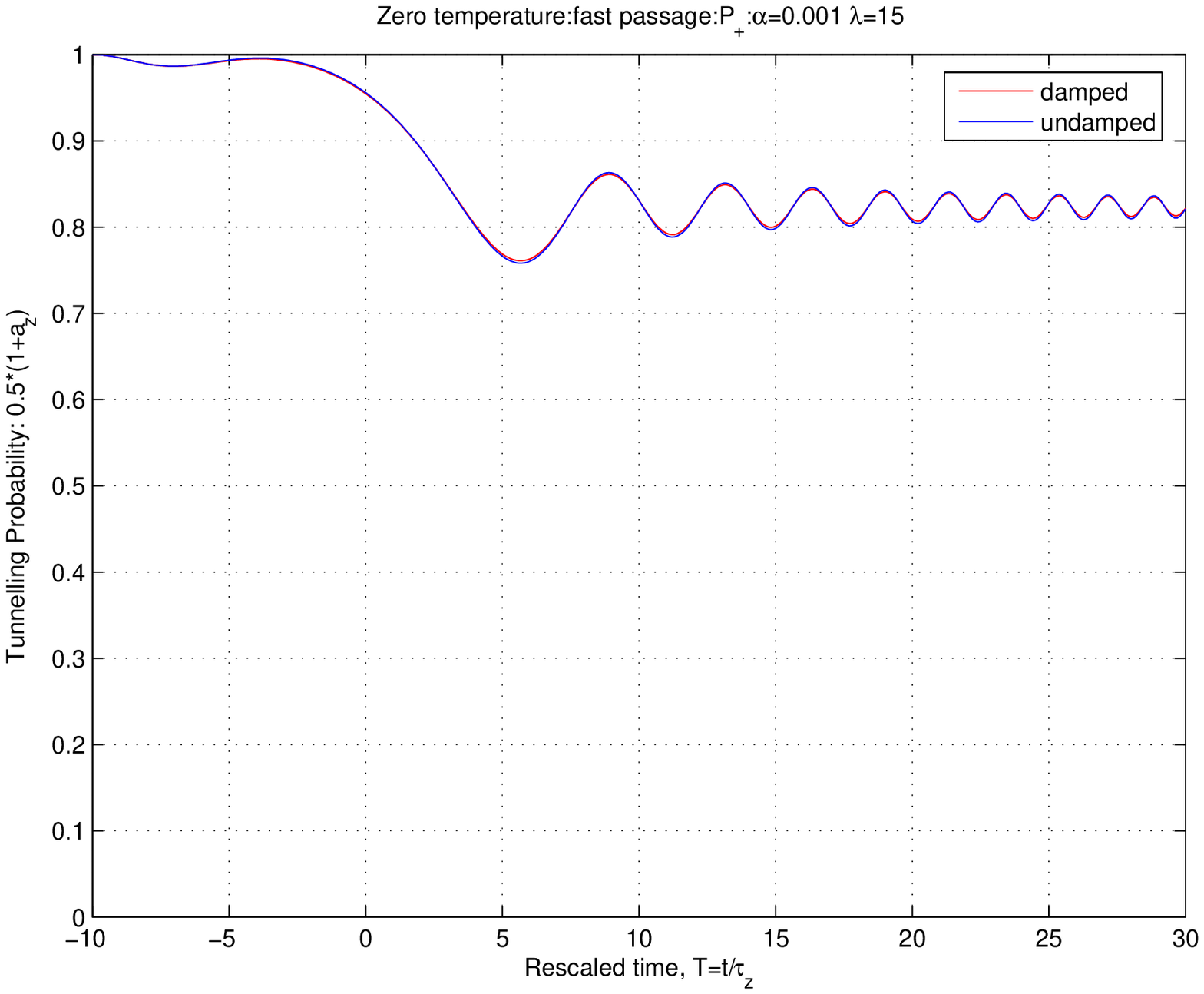}
\caption{As Figure 3 but for the fast passage case $\protect\lambda=15$.}
\label{fig:wide}
\end{figure*}

\begin{figure*}[tbp]
\includegraphics{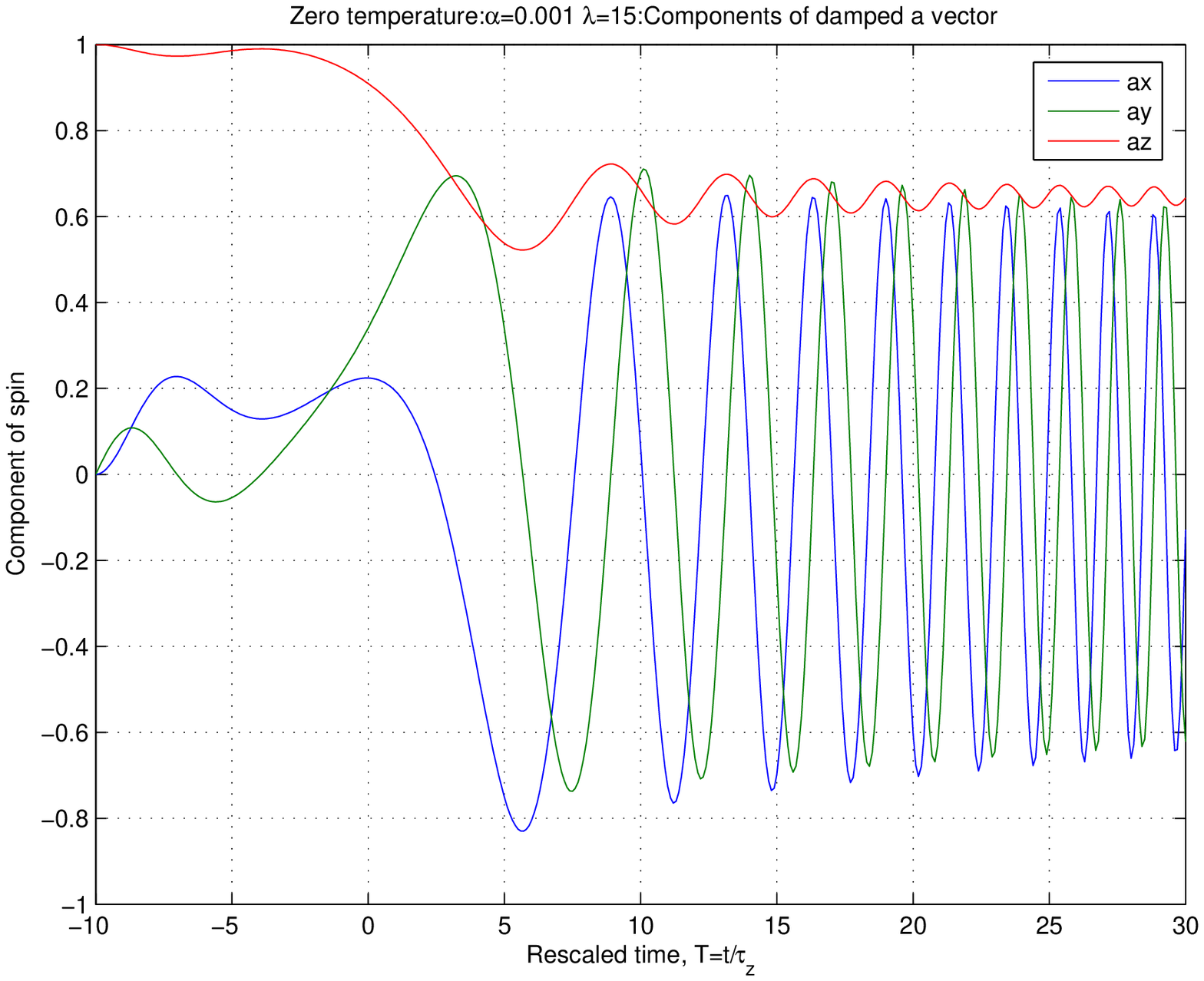}
\caption{As Figure 4 but for the fast passage case $\protect\lambda=15$. }
\label{fig:wide}
\end{figure*}

\begin{figure*}[tbp]
\includegraphics{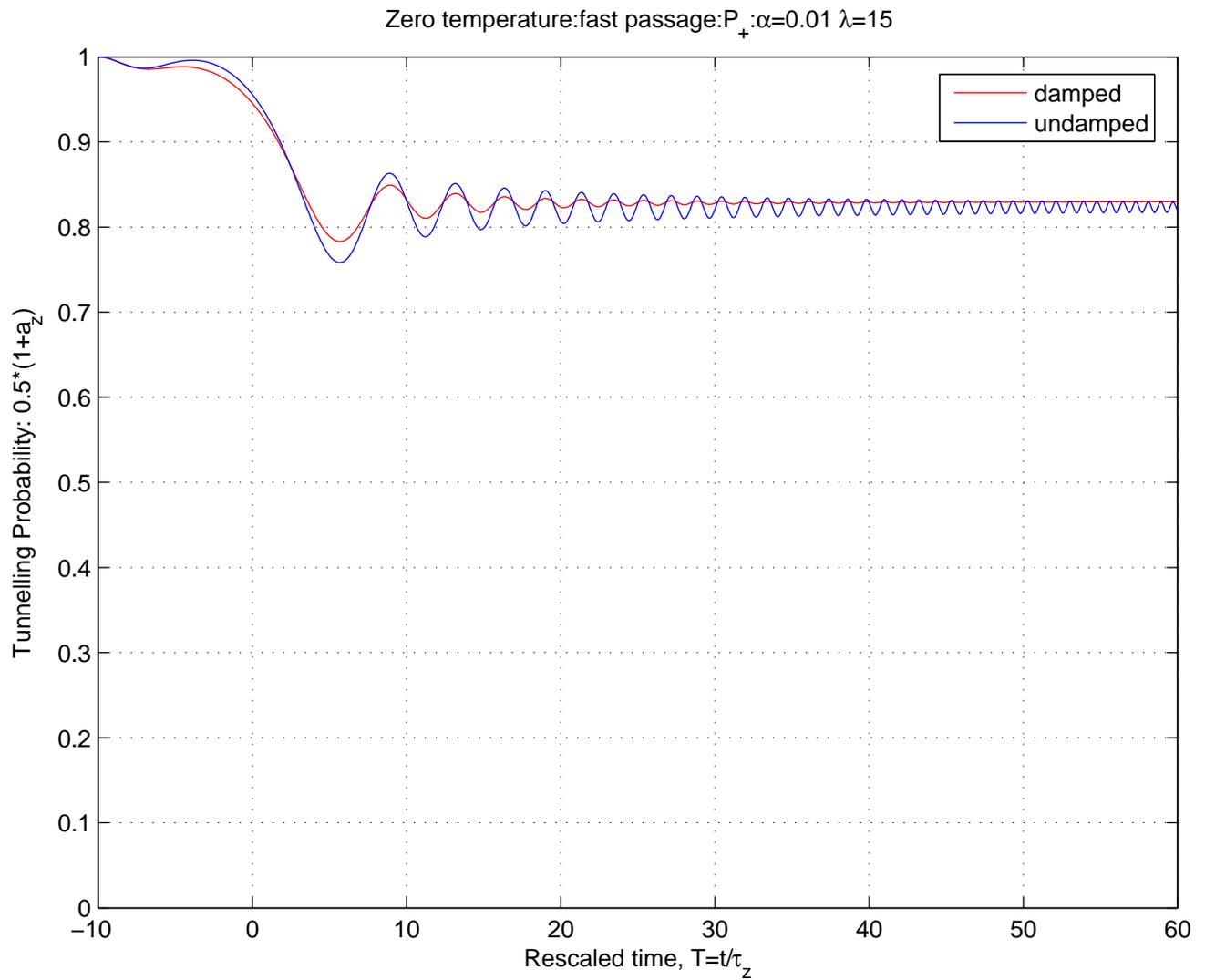}
\caption{As Figure 5 but for stronger coupling to the bath $\protect\alpha%
=0.01$. At this level of coupling the damping of quantum oscillations is
very noticeable, although it does not alter the final tunnelling
probability. }
\label{fig:wide}
\end{figure*}

\begin{figure*}[tbp]
\includegraphics{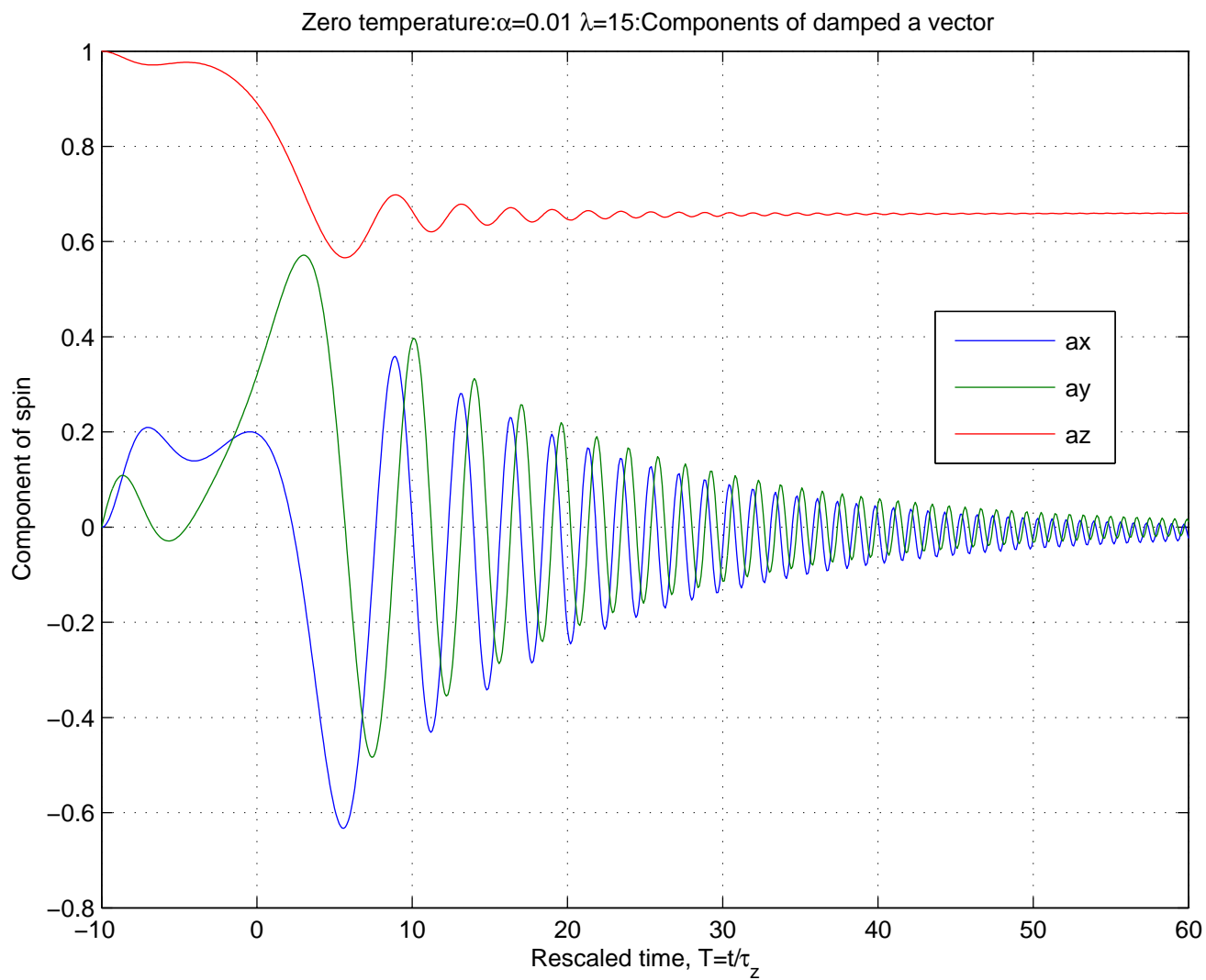}
\caption{As Figure 6 but for stronger coupling to the bath $\protect\alpha%
=0.01$.}
\label{fig:wide}
\end{figure*}

\newpage

\end{document}